\documentclass[preprint]{elsarticle}

\pdfoutput=1 
\usepackage[usenames,dvipsnames]{xcolor}
\usepackage{tikz}
\usepackage{wrapfig}
\usepackage{subfigure}
\usepackage{fixltx2e}
\usepackage{amsmath}
\usepackage{amssymb}
\usepackage{amsfonts}
\usepackage{latexsym}
\usepackage{xspace}
\usepackage{rotating}
\usepackage{enumerate}
\usepackage{paralist}
\usepackage{pgfplots}
\usepackage{booktabs}
\usepackage{enumitem}
\usepackage{etoolbox}
\usepackage{nicefrac}
\usepackage[protrusion=true,expansion=true]{microtype}

\usepackage{float}
\usepackage{pifont}
\usepackage{todonotes}
\usepackage{tablefootnote}
\usepackage{multirow}
\usepackage{hyperref}

\usetikzlibrary{patterns,arrows,calc,shapes,shadows,decorations.pathmorphing,decorations.pathreplacing,automata,shapes.multipart,positioning,shapes.geometric,fit,circuits,trees,shapes.gates.logic.US,fit}

\usetikzlibrary{automata,fit}
\usetikzlibrary{decorations.pathmorphing}

\tikzstyle{every picture}=[thick]
\tikzstyle{every loop}=[->]
\tikzstyle{every scope}=[>=latex]
\tikzstyle{dot}=[circle,thick,minimum size=0.5mm,fill=black, inner sep=1pt]
\tikzstyle{every state}=[draw=black,line width=.5pt,fill=white,minimum size=10pt,initial text=]

\pgfplotscreateplotcyclelist{mycolor}{%
blue,every mark/.append style={fill=blue!80!black},mark=*, densely dashed\\%
violet,every mark/.append style={fill=violet!80!black},mark=square*, densely dashed\\%
BurntOrange!60!black,every mark/.append style={fill=BurntOrange!80!black},mark=otimes*\\%
red,mark=diamond*,every mark/.append style={fill=red!80!black}\\%
}

\input{dft-tikz}

\newcommand{\ie}{i.e., }
\newcommand{\eg}{e.g., }
\newcommand{\wrt}{w.r.t.\ }

\newcommand{\blocks}{\ensuremath{\textsf{Blc}}}
\newcommand{\channels}{\ensuremath{\rhd}}
\newcommand{\inchannels}[1]{\ensuremath{\textsf{in}_{#1}}}
\newcommand{\outchannels}[1]{\ensuremath{\textsf{out}_{#1}}}
\newcommand{\buses}{\ensuremath{\textsf{Bus}}}
\newcommand{\hardwareplatforms}{\ensuremath{H}}
\newcommand{\hardwareassignment}{\ensuremath{h}}
\newcommand{\blockdiagram}{\ensuremath{\mathcal{D}}}
\newcommand{\toplevel}[1]{\ensuremath{\textsf{TLE}(#1)}}
\newcommand{\internal}[1]{\ensuremath{\textsf{internal}_{#1}}}

\newcommand{\storm}{\textsc{Storm}}

\newcommand{\AND}{\textsf{AND}\xspace}
\newcommand{\OR}{\textsf{OR}\xspace}
\newcommand{\PAND}{\textsf{PAND}\xspace}

\newcommand{\VOT}[1]{\textsf{VOT}\textsubscript{#1}\xspace}
\newcommand{\SEQ}{\textsf{SEQ}\xspace}
\newcommand{\SPARE}{\textsf{SPARE}\xspace}
\newcommand{\FDEP}{\textsf{FDEP}\xspace}
\newcommand{\ADEP}{\textsf{ADEP}\xspace}

\newcommand{\BE}{\textsf{BE}\xspace}

\newcommand{\RRplus}{\ensuremath{\mathbb{R}_{>0}}}

\newcommand{\ctmc}{\ensuremath{\mathcal{C}}}
\newcommand{\states}{\ensuremath{S}}
\newcommand{\probability}{\ensuremath{P}}
\newcommand{\rate}{\ensuremath{R}}
\newcommand{\atomiclabel}{\ensuremath{L}}
\newcommand{\finally}[1]{\ensuremath{\lozenge\,#1}}
\newcommand{\boundedfinally}[2]{\ensuremath{\lozenge^{\leq #1}\,#2}}
\newcommand{\until}[2]{\ensuremath{#1\,\mathbf{\mathsf{U}}\,#2}}
\newcommand{\boundeduntil}[3]{\ensuremath{#1\,\mathbf{\mathsf{U}}^{\leq #2}\,#3}}

\newcommand{\bad}{\ensuremath{bad}\xspace}
\newcommand{\target}{\ensuremath{target}\xspace}

\newcommand{\PROB}{\ensuremath{\mathsf{P}}}
\newcommand{\ET}{\ensuremath{\mathsf{ET}}}
\newcommand{\AFH}{AFH\xspace}
\newcommand{\MTTF}{MTTF\xspace}
\newcommand{\FFA}{FFA\xspace}
\newcommand{\FWD}{FWD\xspace}
\newcommand{\MTDF}{MTDF\xspace}
\newcommand{\MDR}{MDR\xspace}
\newcommand{\FLOD}{FLOD\xspace}
\newcommand{\SILFO}{SILFO\xspace}

\newdefinition{definition}{Definition}
\newtheorem{example}{Example}

\definecolor{lightblue}{RGB}{224,224,255}
\definecolor{lightred}{RGB}{255,224,224}
\definecolor{lightgreen}{RGB}{224,255,224}
\definecolor{lightyellow}{RGB}{255,255,224}
\definecolor{lightpurple}{RGB}{255,224,255}
\definecolor{darkerred}{RGB}{64,0,0}
\definecolor{darkred}{RGB}{128,0,0}
\definecolor{darkblue}{RGB}{0,0,128}
\definecolor{darkgreen}{RGB}{0,128,0}
\definecolor{darkpurple}{RGB}{128,0,128}

\journal{Reliability Engineering \& System Safety}

\begin{document}

\begin{frontmatter}

\title{Safety Analysis for Vehicle Guidance Systems\\with Dynamic Fault Trees\tnoteref{copyright}\tnoteref{funding}}
\tnotetext[copyright]{\copyright\,2019. This preprint version is made available under the CC-BY-NC-ND 4.0 license \url{http://creativecommons.org/licenses/by-nc-nd/4.0/}. The full paper is available in RESS \href{https://doi.org/10.1016/j.ress.2019.02.005}{10.1016/j.ress.2019.02.005}.}
\tnotetext[funding]{Funding: This work was supported by the CDZ project CAP and the DFG RTG 2236 ``UnRAVeL''.}

\author[bmw]{Majdi Ghadhab}
\author[rwth]{Sebastian Junges}
\author[rwth]{Joost-Pieter Katoen}
\author[bmw]{Matthias Kuntz}
\author[rwth]{Matthias Volk}
\ead{matthias.volk@cs.rwth-aachen.de}

\address[bmw]{BMW AG, Munich, Germany}
\address[rwth]{RWTH Aachen University, Aachen, Germany}

\newlength{\miniskip}
\setlength{\miniskip}{2pt plus 1pt minus 1pt}
\renewcommand{\subsubsection}[1]{\smallskip\noindent\textbf{#1.}}
\renewcommand{\paragraph}[1]{\smallskip\noindent\textit{#1}}

\newcommand{\dftscale}{0.6}

\begin{abstract}
This paper considers the design-phase safety analysis of vehicle guidance systems. 
The proposed approach constructs dynamic fault trees (DFTs) to model a variety of safety concepts and E/E architectures for drive automation. 
The fault trees can be used to evaluate various quantitative measures by means of model checking. 
The approach is accompanied by a large-scale evaluation: The resulting DFTs with up to 300 elements constitute larger-than-before DFTs, yet the concepts and architectures can be evaluated in a matter of minutes.
\end{abstract}

\end{frontmatter}


\section{Introduction}
\label{sec:intro}

\paragraph{Motivation}
Cars are nowadays equipped with functions, often realised in software, to e.g., improve driving comfort and driving assistance (with a tendency towards autonomous driving).
These functions impose high demands on the required functional safety. 
ISO 26262~\cite{iso26262} is the basic norm for developing safety-critical functions in the automotive setting.
It enables car manufacturers to develop safety-critical devices---in the sense that malfunctioning can harm persons---according to an agreed technical state-of-the-art. 
The safety-criticality is technically measured in terms of the so-called Automotive Safety Integrity Level (ASIL). This level takes into account driving situations, failure occurrence, the possible resulting physical harm, and the controllability of the malfunctioning by the driver. The result is classified from QM (no special safety measures required) up to the most stringent level ASIL D (with ASIL A, B, C in between).
To meet the functional safety requirements, it is crucial to execute the software functions with a sufficiently low probability of undetected dangerous hardware failures.
This paper considers the design-phase safety analysis of the \emph{vehicle guidance} system, a key functional block of a vehicle with a high safety integrity level (ASIL D, i.e., allowing not more than 10$^{{-}8}$ residual hardware failures per hour). 
The key point of our approach is to: (1) manually construct dynamic fault trees~\cite{DBB90} (DFTs) from industrial system descriptions and combine them (in an automated manner) with hardware failure models for several partitionings of functions on hardware, and (2) analyse the resulting overall DFTs by means of probabilistic model checking~\cite{Kat16,DBLP:conf/lics/Kwiatkowska03,DBLP:series/natosec/Baier16}. 

\begin{figure}[t]
\centering
\scalebox{0.8}{
\begin{tikzpicture}
	\node[rectangle, draw, minimum height=1cm, minimum width=3cm, align=center] at (2.4,3.45) (fbd)  {functional\\block diagram};

	\node[rectangle, draw, minimum width=0.8cm, minimum height=3cm, anchor=center] at (15.2, 2.4) (analyse) {\rotatebox{270}{analyse, cf. Fig.~\ref{fig:dft_toolchain}}};

	\draw (5.9,3.5) node[anchor=north]{}
  -- (9.1,3.5) node[anchor=north]{}
  -- (7.5,3.95) node[anchor=north]{system}
  -- cycle;
  \draw(6.5, 3.5) node[anchor=north,yshift=-7pt] {$B_1$}
  -- (5.8, 2.8)
  -- (7.2, 2.8) -- cycle;
    \draw(8.5, 3.5) node[anchor=north,yshift=-7pt] {$B_4$}
  -- (7.8, 2.8)
  -- (9.2, 2.8) -- cycle;
  \node[] at (7.5, 3) {$\hdots$};

   	\node[rectangle, draw, minimum height=0.7cm, minimum width=3cm, below=0.22cm of fbd] (eea) {E/E architecture};
	\node[rectangle, draw, minimum height=1cm, below=0.1cm of eea, align=center, minimum width=3cm] (hwft) {hardware\\(+fault trees)};
	\node[rectangle, draw, minimum height=1.1cm, minimum width=3.3cm, right=1.8cm of eea, align=center, yshift=-0.2cm] (hwa) {hardware assignment};
	\draw (11,3.5) node[anchor=north]{}
  -- (14.2,3.5) node[anchor=north]{}
  -- (12.6,3.95) node[anchor=north]{system}
  -- cycle;
  \draw(11.6, 3.5) node[anchor=north,yshift=-7pt] {$B_1$}
  -- (10.9, 2.8)
  -- (12.3, 2.8) -- cycle;
    \draw(13.6, 3.5) node[anchor=north,yshift=-7pt] {$B_4$}
  -- (12.9, 2.8)
  -- (14.3, 2.8) -- cycle;
  \node[] at (12.6, 3) {$\hdots$};

   \draw(11.3, 1.7) node[anchor=north,yshift=-7pt] {$\hardwareplatforms_1$}
  -- (10.7, 0.9)
  -- (11.9, 0.9) -- cycle;
   \draw(12.6, 1.7) node[anchor=north,yshift=-7pt] {$\hardwareplatforms_2$}
  -- (12, 0.9)
  -- (13.2, 0.9) -- cycle;
   \draw(13.9, 1.7) node[anchor=north,yshift=-7pt] {$\buses_1$}
  -- (13.3, 0.9)
  -- (14.5, 0.9) -- cycle;

   \draw[dashed] (11.3, 1.7) -- (11.2, 2.8);
   \draw[dashed] (12.6, 1.7) -- (13.2, 2.8);
   \draw[dashed] (13.9, 1.7) -- (11.8, 2.8);
   \draw[dashed] (13.9, 1.7) -- (13.8, 2.8);

   \draw[->, very thick] (3.9, 3.5) -- node[above] {Sect.~\ref{Subsec:FBD2FFT}} (5.8, 3.5);
   \draw[->, very thick] (3.9, 1.1) -- node[above] {Sect.~\ref{Subsec:FT_hardware}} (10, 1.1);
   \draw[->, very thick] (3.9, 2.4) -- node[above] {} (5.7, 2.4);
   \draw[->, very thick] (3.9, 1.7) -- node[above] {Sect.~\ref{Subsec:Hardware_assignment}} (5.7, 1.7);
   \draw[->, very thick] (9.2, 2.1) -- node[above] {Sect.~\ref{Subsec:Constructing_complete_FT}} (11.2, 2.1);
   \draw[->, very thick] (14.1, 2.4) -- node[above] {} (14.8, 2.4);
  
\end{tikzpicture}
}
\caption{Overview of the model-based safety approach}
\label{fig:overview}
\end{figure}
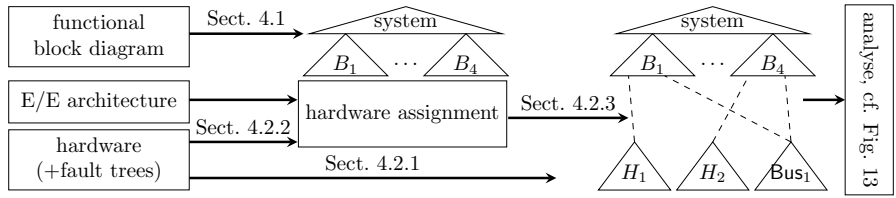

\paragraph{A model-based approach}
Fig.~\ref{fig:overview} summarises the approach, in relation to the structure of this paper.
The failure behaviour of the functional architecture, given as a functional block diagram (FBD), is expressed as a \emph{two-level} DFT: the upper level models a \emph{system failure} in terms of block failures $B_i$ while the lower level models the causes of \emph{block failures} $B_i$.
The use of fault trees is natural: They are a well-known model in reliability engineering. No familiarity with additional formalisms is required. Fault trees for hardware components are typically provided by manufacturers. Failures in function blocks can easily be described by fault trees.
The use of DFTs rather than static fault trees allows to model warm and cold redundancies, spare components, and state-dependent faults; cf.~\cite{RS15}.
Each functional block is assigned to a hardware platform for which (by assumption) a provided DFT $\hardwareplatforms_i$ models its failure behaviour.
Depending on the partitioning, the communication goes via different fallible buses that are also modelled by DFTs $\buses_i$.
From the partitioning, and the DFTs of the hardware and the functional level, an overall DFT is constructed (in an automated manner) consisting of three layers: (1) the system layer; (2) the block layer; and (3) the hardware layer. Details are discussed in Sect.~\ref{sec:modelling}.

\paragraph{Analysis}
We exploit probabilistic model checking (PMC)~\cite{Kat16,DBLP:conf/lics/Kwiatkowska03,DBLP:series/natosec/Baier16} to analyse the DFT of the overall vehicle guidance system.
PMC can be used as a black-box algorithm---no expertise in PMC is needed to understand its outcomes---and supports \emph{various metrics} that go beyond reliability and MTTF~\cite{VJK17}.
While they are all expressible by \emph{a combination of PMC queries}, the number of queries is prohibitively large for some measures relevant for the safety analysis of highly automated cars.
Therefore, we developed dedicated algorithms to compute these measures within the probabilistic model checker \storm~\cite{DJKV17}, by reusing building blocks for standard PMC queries.
In contrast to simulation and statistical model checking \cite{DBLP:conf/rv/LegayDB10,DBLP:journals/tomacs/AghaP18}, where results are obtained with a given statistical confidence, PMC provides \emph{hard guarantees} that the safety objectives are met.
These guarantees are important as ISO 26262 requires that
``metrics are verifiable and precise enough to differentiate between different architectures''\cite[5:8.2]{iso26262}.

Whereas most ISO 26262-based analyses focus on single and dual-point failures, PMC naturally supports the analysis of \emph{multi-point failures} of the vehicle guidance system's DFT.
Consideration of multi-point failures is highly relevant, as ``it is necessary to consider multiple-point failures of a higher order than two in the analysis when the technical safety concept is based on redundant safety mechanisms.''\cite[5:9.4.3.2]{iso26262}.
To limit the computation time, we extend \storm{'s} capabilities for approximative computation: Instead of a precise value, we compute sound upper and lower bounds for the measures.

\paragraph{Contributions} 
The main contribution of this paper is two-fold: We report on the usage of dynamic fault trees for safety analysis in a potential automotive setting. 
While standard fault tree analysis is part of the ISO 26262, the usage of DFTs in this field is new. 
\emph{This paper shows how the additional features offered by DFTs help to create faithful models of the considered scenarios.} These models are then used to analyse the given scenarios.
To increase the applicability of DFTs as a method for probabilistic safety assessment in an industrial setting, we give concrete building blocks to work with, e.g.~redundancy and faults covered by fallible safety mechanisms. 

A clear benefit of the usage of DFTs is that all these methods are integrated in existing off-the-shelf analysis tools, which provide sound error bounds. 
The usage of DFTs reduces the amount of domain-specific knowledge in the analysis, and thus supports a more model-oriented approach. 
\emph{In this paper, we take this model-oriented approach to investigate the effect of different hardware partitioning on a range of metrics.}
The generated DFTs are to the best of our knowledge the largest real-life ones in the literature---larger trees have only been artificially created for scalability purposes~\cite{JGKRS17}.
Notably, this paper is the first to consider model-checking based approaches for DFT analysis on real-life case studies. 

A short version of this paper was presented in~\cite{DBLP:conf/safecomp/GhadhabJKKV17}.
The three major extensions to the earlier paper are:
 (1) A more detailed step-by-step description of the methodology. 
 (2) New and faster algorithms to compute a number of safety measures, and a more thorough explanation of the algorithms involved. 
 (3) More thorough experiments and a detailed analysis of the results.

\paragraph{Remark}
This work has been carried out in cooperation with BMW AG. 
The proposed concepts and architectures are exemplary for real-life systems.
No implication on actual safety concepts or E/E architectures implemented by BMW AG can be derived from these examples. 
The same remark applies on any quantity (failure rates, obtained metrics, ...) presented in this paper.

\section{Vehicle Guidance}
\label{sec:case}

The most challenging safety topic in the automotive industry is currently the driving automation, where the driving responsibility is moving partly or even entirely from the driver to the embedded vehicle intelligence. Rising liability questions make it crucial to develop functional safety concepts adequately to the intended automation level and to provide evidence regarding the availability and the reliability of these concepts.

\tikzstyle{ablock}=[rectangle, draw, inner sep=1pt, minimum width=1.05cm, minimum height=0.6cm]
\tikzstyle{fblock}=[rectangle, draw, inner sep=1pt, minimum width=1.05cm, minimum height=0.8cm]

\tikzstyle{actuator}=[rectangle, draw, inner sep=1pt, scale=0.95, minimum width=0.6cm, minimum height=0.4cm]
\tikzstyle{sensor}=[rectangle, draw, inner sep=1pt, scale=0.95, minimum width=0.6cm, minimum height=0.4cm]

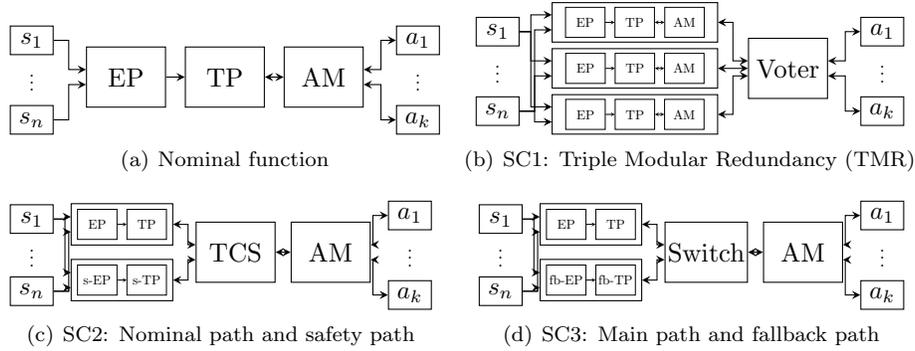
\begin{figure}[t]
\centering
\subfigure[Nominal function]{
\begin{tikzpicture}
	\node[fblock] (EP) {EP};
	\node[fblock, right=0.25cm of EP] (TP) {TP};
	\node[fblock, right=0.25cm of TP] (AM) {AM};
	\node[left=0.6cm of EP, scale=0.7] (sx) {$\vdots$};
	\node[sensor, above=0.02cm of sx] (s1) {$s_1$};
	\node[sensor, below=0.1cm of sx] (sn) {$s_n$};
	\node[right=0.6cm of AM, scale=0.7] (ax) {$\vdots$};
	\node[actuator, above=0.02cm of ax] (a1) {$a_1$};
	\node[actuator, below=0.1cm of ax] (an) {$a_k$};
	
	\draw[->] (s1.east) -- +(0.2,0) |- ([yshift=0.1cm]EP.west);
	\draw[->] (sn.east) -- +(0.2,0) |- ([yshift=-0.1cm]EP.west);
	\draw[->] (EP) -- (TP);
	\draw[<->] (TP) -- (AM);
	\draw[<->] (a1.west) -- +(-0.2,0) |- ([yshift=0.1cm]AM.east);
	\draw[<->] (an.west) -- +(-0.2,0) |- ([yshift=-0.1cm]AM.east);
\end{tikzpicture}
\label{Fig:scen_nom}
}\hfill
\subfigure[SC1: Triple Modular Redundancy (TMR)]{
\begin{tikzpicture}

	\node[rectangle, minimum height=1.05cm, draw, scale=0.5, minimum width=4.4cm] (p2) {\pgftext{\begin{tikzpicture}\node[fblock] (EP) {EP};
	\node[fblock, right=0.25cm of EP] (TP) {TP};
	\node[fblock, right=0.25cm of TP] (AM) {AM};
	\draw[->] (EP) -- (TP);
	\draw[<->] (TP) -- (AM);\end{tikzpicture}}};
	\node[rectangle, minimum height=1.05cm, draw, scale=0.5, minimum width=4.4cm, above=0.08cm of p2] (p1) {\pgftext{\begin{tikzpicture}\node[fblock] (EP) {EP};
	\node[fblock, right=0.25cm of EP] (TP) {TP};
	\node[fblock, right=0.26cm of TP] (AM) {AM};
	\draw[->] (EP) -- (TP);
	\draw[<->] (TP) -- (AM);\end{tikzpicture}}};
	\node[rectangle, minimum height=1.05cm, draw, scale=0.5, minimum width=4.4cm, below=0.08cm of p2] (p3) {\pgftext{\begin{tikzpicture}\node[fblock] (EP) {EP};
	\node[fblock, right=0.25cm of EP] (TP) {TP};
	\node[fblock, right=0.25cm of TP] (AM) {AM};
	\draw[->] (EP) -- (TP);
	\draw[<->] (TP) -- (AM);\end{tikzpicture}}};
	
	\node[fblock, right=0.4cm of p2] (VOT) {Voter};
	
	\node[left=0.6cm of p2, scale=0.7] (sx) {$\vdots$};
	\node[sensor, above=0.02cm of sx] (s1) {$s_1$};
	\node[sensor, below=0.1cm of sx] (sn) {$s_n$};
	\node[right=0.6cm of VOT, scale=0.7] (ax) {$\vdots$};
	\node[actuator, above=0.02cm of ax] (a1) {$a_1$};
	\node[actuator, below=0.1cm of ax] (an) {$a_k$};
	
	\draw[->] (s1.east) -- +(0.15,0) |- ([yshift=0.1cm]p2.west);
	\draw[->] (sn.east) -- +(0.2,0) |- ([yshift=-0.1cm]p2.west);
	\draw[->] (s1.east) -- +(0.15,0) |- ([yshift=0.1cm]p1.west);
	\draw[->] (sn.east) -- +(0.2,0) |- ([yshift=-0.1cm]p1.west);
	\draw[->] (s1.east) -- +(0.15,0) |- ([yshift=0.1cm]p3.west);
	\draw[->] (sn.east) -- +(0.2,0) |- ([yshift=-0.1cm]p3.west);
	\draw[<->] (a1.west) -- +(-0.2,0) |- ([yshift=0.1cm]VOT.east);
	\draw[<->] (an.west) -- +(-0.2,0) |- ([yshift=-0.1cm]VOT.east);
	\draw[<->] (p1.east) -- +(0.2,0) |- ([yshift=0.1cm]VOT.west);
	\draw[<->] (p2.east) -- +(0.2,0) |- (VOT.west);
	\draw[<->] (p3.east) -- +(0.2,0) |- ([yshift=-0.1cm]VOT.west);
\end{tikzpicture}
\label{Fig:scen_tmr}
}	
\subfigure[SC2: Nominal path and safety path]{
\begin{tikzpicture}
    
    \node[rectangle, minimum width=2.7cm, scale=0.5] (p2) {};
	\node[rectangle, minimum height=1.1cm, draw, scale=0.5, minimum width=2.7cm, above=0.03cm of p2] (p1) {\pgftext{\begin{tikzpicture}\node[fblock] (EP) {EP};
	\node[fblock, right=0.25cm of EP] (TP) {TP};
	\draw[->] (EP) -- (TP);\end{tikzpicture}}};
	\node[rectangle, minimum height=1.1cm, draw, scale=0.5, minimum width=2.7cm, below=0.03cm of p2] (p3) {\pgftext{\begin{tikzpicture}\node[fblock] (EP) {s-EP};
	\node[fblock, right=0.25cm of EP] (TP) {s-TP};
	\draw[->] (EP) -- (TP);\end{tikzpicture}}};
	
	\node[fblock, right=0.3cm of p2] (VOT) {TCS};
	\node[fblock, right=0.2cm of VOT] (AM) {AM};
	
	\node[left=0.4cm of p2, scale=0.7] (sx) {$\vdots$};
	\node[sensor, above=0.02cm of sx, yshift=-0.05cm] (s1) {$s_1$};
	\node[sensor, below=0.05cm of sx] (sn) {$s_n$};
	\node[right=0.4cm of AM, scale=0.7] (ax) {$\vdots$};
	\node[actuator, above=0.02cm of ax] (a1) {$a_1$};
	\node[actuator, below=0.1cm of ax] (an) {$a_k$};

	\draw[<->] (VOT) -- (AM);
	
	\draw[->] (s1.east) -- +(0.15,0) |- ([yshift=0.1cm]p1.west);
	\draw[->] (sn.east) -- +(0.2,0) |- ([yshift=-0.1cm]p1.west);
	\draw[->] (s1.east) -- +(0.15,0) |- ([yshift=0.1cm]p3.west);
	\draw[->] (sn.east) -- +(0.2,0) |- ([yshift=-0.1cm]p3.west);
	\draw[<->] (a1.west) -- +(-0.2,0) |- ([yshift=0.1cm]AM.east);
	\draw[<->] (an.west) -- +(-0.2,0) |- ([yshift=-0.1cm]AM.east);
	\draw[<->] (p1.east) -- +(0.2,0) |- ([yshift=0.1cm]VOT.west);
	\draw[<->] (p3.east) -- +(0.2,0) |- ([yshift=-0.1cm]VOT.west);
\end{tikzpicture}
\label{Fig:scen_npsp}
}\hfill
\subfigure[SC3: Main path and fallback path]{
\begin{tikzpicture}
    
    \node[rectangle, minimum width=2.7cm, scale=0.5] (p2) {};
	\node[rectangle, minimum height=1.1cm, draw, scale=0.5, minimum width=2.7cm, above=0.03cm of p2] (p1) {\pgftext{\begin{tikzpicture}\node[fblock] (EP) {EP};
	\node[fblock, right=0.25cm of EP] (TP) {TP};
	\draw[->] (EP) -- (TP);\end{tikzpicture}}};
	\node[rectangle, minimum height=1.1cm, draw, scale=0.5, minimum width=2.7cm, below=0.03cm of p2] (p3) {\pgftext{\begin{tikzpicture}\node[fblock] (EP) {fb-EP};
	\node[fblock, right=0.25cm of EP] (TP) {fb-TP};
	\draw[->] (EP) -- (TP);\end{tikzpicture}}};
	
	\node[fblock, right=0.3cm of p2] (VOT) {Switch};
	\node[fblock, right=0.2cm of VOT] (AM) {AM};
	
	\node[left=0.4cm of p2, scale=0.7] (sx) {$\vdots$};
	\node[sensor, above=0.02cm of sx, yshift=-0.05cm] (s1) {$s_1$};
	\node[sensor, below=0.05cm of sx] (sn) {$s_n$};
	\node[right=0.4cm of AM, scale=0.7] (ax) {$\vdots$};
	\node[actuator, above=0.02cm of ax] (a1) {$a_1$};
	\node[actuator, below=0.1cm of ax] (an) {$a_k$};
	
	\draw[<->] (VOT) -- (AM);
	\draw[->] (s1.east) -- +(0.15,0) |- ([yshift=0.1cm]p1.west);
	\draw[->] (sn.east) -- +(0.2,0) |- ([yshift=-0.1cm]p1.west);
	\draw[->] (s1.east) -- +(0.15,0) |- ([yshift=0.1cm]p3.west);
	\draw[->] (sn.east) -- +(0.2,0) |- ([yshift=-0.1cm]p3.west);
	\draw[<->] (a1.west) -- +(-0.2,0) |- ([yshift=0.1cm]AM.east);
	\draw[<->] (an.west) -- +(-0.2,0) |- ([yshift=-0.1cm]AM.east);
	\draw[<->] (p1.east) -- +(0.2,0) |- ([yshift=0.1cm]VOT.west);
	\draw[<->] (p3.east) -- +(0.2,0) |- ([yshift=-0.1cm]VOT.west);
\end{tikzpicture}
\label{Fig:scen_mpfp}
}		
\caption{Different functional block diagrams for vehicle guidance}
\label{Fig:nominal}
\end{figure}

\subsection{Scenario}
As a real-life case study from the automotive domain, we consider the functional block diagram (FBD) in Fig.~\ref{Fig:scen_nom} representing the skeletal structure of automated driving. 
Data collected from different sensors (cameras, radars, ultrasonic, etc.) are synthesised and fused to generate a model of the current driving situation in the Environment Perception (EP) functional block. 
This model is used by the Trajectory Planning (TP) functional block to build a driving path with respect to the current driving situation and the intended trip. 
The Actuator Management (AM) functional block ensures the control of the different actuators (powertrain, brakes, etc.) following the calculated driving path. Thus, the blocks in the FBD fulfil tasks:
The tasks are realised by (potentially redundant) functional blocks, connected by lines to depict dataflow.
We like to stress that these diagrams are \emph{not} reliability block diagrams in which the system is operational as long as a path through operational blocks exist. 

\subsection{Modelling Safety Concepts}

\label{Sec:VehicleGuidance_caee}
\subsubsection{Technical safety concepts}
Based on the criticality of the vehicle guidance function, especially when the driver is out-of-the-loop, ASIL D, the highest level, applies to the safety goal of following a safe trajectory. 
According to the automation level, the vehicle guidance function \emph{must} be designed as \emph{fail-operational}, i.e., the system should safely continue to operate for a certain time after a failure of one of its components.
Different design patterns have been developed and implemented in safety-critical systems with fail-operational behaviour and high safety levels, cf. e.g.~\cite{ASK09}.  
The variety of possibilities is illustrated by the following three concepts:
\begin{enumerate}[label={\textbf{SC\arabic*} -}, wide, labelwidth=!, labelindent=0pt]
\item Triple Modular Redundancy (TMR), Fig.~\ref{Fig:scen_tmr}: The nominal function for vehicle guidance is replicated into three paths each fulfilling ASIL B. A Voter, fulfilling ASIL D, ensuring that any single incorrect path is eliminated. 
\item Nominal path and safety path, Fig.~\ref{Fig:scen_npsp}: Consists of two different paths, a nominal path (n-Path) and a safety path (s-Path) in hot-standby mode. 
The n-Path provides a full extent trajectory---including comfort functions not necessary for safe operation---with ASIL QM and the s-Path a reduced extent trajectory but with highest safety integrity level ASIL D. 
The reduced extent safety trajectory is generated from a reduced s-EP (safety Environment Perception) and reduced s-TP (safety Trajectory Planning). 
The Trajectory Checking and Selection (TCS) verifies whether the trajectory calculated by the n-Path is within the safe range calculated by the s-Path or not. 
In the case of failure, the s-Path takes over the control and the safe trajectory with reduced extent is followed by the AM. 
In this case, the system is considered to be \emph{degraded}.
\item Main path and fallback path, Fig.~\ref{Fig:scen_mpfp}: Similar to SC2 although the main path (m-Path) is now developed according to ASIL D in order to detect its own hardware failures and signalise them to the Switch. The Switch then commutates the control of the AM to a fallback path (fb-Path), operated in cold-standby, with ASIL D. Upon activation of the fb-Path, the system is considered to be \emph{degraded}. 
\end{enumerate}

\begin{figure}[t]
\centering
	\subfigure[E/E architecture A]{\scalebox{0.9}{
	\begin{tikzpicture}
		\draw[thick] (3, 3) -- (3, 1.1);
		\node[ablock] at (2.3, 2.5) (adas) {ADAS};
		\node[ablock] at (2.3, 1.3) (iecu) {I-ECU};
		\node[ablock] at (3.7, 2.6) (ecu1) {ECU$_1$};
		\node[] at (3.7, 2.1) {$\vdots$};
		\node[ablock] at (3.7, 1.4) (ecu3) {ECU$_k$};
		
		\node[left=0.4cm of adas, scale=0.7] (sx) {$\vdots$};
	\node[sensor, above=0.02cm of sx] (s1) {$s_1$};
	\node[sensor, below=0.1cm of sx] (sn) {$s_n$};
	
	\node[actuator, right=0.2cm of ecu1] (a1) {$a_1$};
	\node[actuator, right=0.2cm of ecu3] (a3) {$a_k$};

	\node[actuator, left=0.3cm of iecu] (ia) {$a_0$};

		\draw[-] (s1.east) -- +(0.1,0) |- ([yshift=0.08cm]adas.west);
		\draw[-] (sn.east) -- +(0.1,0) |- ([yshift=-0.08cm]adas.west);

		\node at (3, 0.8) (bus_label) {Bus};
		\draw (adas.east) -- (3, 2.5);
		\draw (ecu1.west) -- (3, 2.6);
		\draw (ecu3.west) -- (3, 1.4);
		 \draw (iecu.east) -- (3, 1.3);
		 \draw (ecu1) -- (a1);
		 \draw (ecu3) -- (a3);
		 \draw (iecu) -- (ia);
	\end{tikzpicture}}
	\label{fig:ee_nominal}
	}\hfill
	\subfigure[E/E architecture B]{\scalebox{0.9}{
	\begin{tikzpicture}
		\draw[thick] (3, 3.4) -- (3, 1.1);
		\draw[thick] (1.6, 3.4) -- (1.6, 1.1);

		\node[ablock] at (2.3, 3.0) (adas) {ADAS$_1$};
		\node[ablock] at (2.3, 2.25) (adas2) {ADAS$_2$};
		\node[ablock] at (2.3, 1.5) (adas3) {ADAS$_3$};
		\node[ablock] at (3.7, 3.3) (iecu) {I-ECU};
		\node[ablock] at (3.7, 2.6) (ecu1) {ECU$_1$};
		\node[] at (3.7, 2.1) {$\vdots$};
		\node[ablock] at (3.7, 1.4) (ecu3) {ECU$_k$};
		
		\node[left=0.4cm of adas2, scale=0.7] (sx) {$\vdots$};
	\node[sensor, above=0.02cm of sx] (s1) {$s_1$};
	\node[sensor, below=0.1cm of sx] (sn) {$s_n$};
	
	\node[actuator, right=0.2cm of ecu1] (a1) {$a_1$};
	\node[actuator, right=0.2cm of ecu3] (a3) {$a_k$};

	\node[actuator, right=0.2cm of iecu] (ia) {$a_0$};

		\draw[-] (s1.east) -| (1.6, 2);
		\draw[-] (sn.east) -| (1.6, 2);

		\node at (3, 0.9) (bus_label) {Bus};
		\node at (1.5, 0.9) (bus_label) {Bus};
		\draw (adas.east) -- (3, 3.0);
		\draw (adas2.east) -- (3, 2.25);
		\draw (adas3.east) -- (3, 1.5);
		\draw (adas2.west) -- (1.6, 2.25);
		\draw (adas.west) -- (1.6, 3.0);
		\draw (adas3.west) -- (1.6, 1.5);
		\draw (ecu1.west) -- (3, 2.6);
		\draw (ecu3.west) -- (3, 1.4);
		 \draw (iecu.west) -- (3, 3.3);
		 \draw (ecu1) -- (a1);
		 \draw (ecu3) -- (a3);
		 \draw (iecu) -- (ia);
	\end{tikzpicture}}
	\label{fig:ee_tmr}
	}\hfill
	\subfigure[E/E architecture C]{\scalebox{0.9}{
	\begin{tikzpicture}
		\draw[thick] (3, 3.4) -- (3, 1.1);
		\draw[thick] (1.6, 3.4) -- (1.6, 1.1);

		\node[ablock] at (2.3, 3.0) (adas) {ADAS$_1$};
		\node[ablock] at (2.3, 2.25) (adas2) {ADAS$^{+}_2$};
		\node[ablock] at (3.7, 3.3) (iecu) {I-ECU};
		\node[ablock] at (3.7, 2.6) (ecu1) {ECU$_1$};
		\node[] at (3.7, 2.1) {$\vdots$};
		\node[ablock] at (3.7, 1.4) (ecu3) {ECU$_k$};
		
		\node[left=0.4cm of adas2, scale=0.7, yshift=0.3cm] (sx) {$\vdots$};
	\node[sensor, above=0.02cm of sx] (s1) {$s_1$};
	\node[sensor, below=0.1cm of sx] (sn) {$s_n$};
	
	\node[actuator, right=0.2cm of ecu1] (a1) {$a_1$};
	\node[actuator, right=0.2cm of ecu3] (a3) {$a_k$};

	\node[actuator, right=0.2cm of iecu] (ia) {$a_0$};

		\draw[-] (s1.east) -| (1.6, 2);
		\draw[-] (sn.east) -| (1.6, 2);

		\node at (3, 0.9) (bus_label) {Bus};
		\node at (1.5, 0.9) (bus_label) {Bus};
		\draw (adas.east) -- (3, 3.0);
		\draw (adas2.east) -- (3, 2.25);
		\draw (adas2.west) -- (1.6, 2.25);
		\draw (adas.west) -- (1.6, 3.0);
		\draw (ecu1.west) -- (3, 2.6);
		\draw (ecu3.west) -- (3, 1.4);
		 \draw (iecu.west) -- (3, 3.3);
		 \draw (ecu1) -- (a1);
		 \draw (ecu3) -- (a3);
		 \draw (iecu) -- (ia);
	\end{tikzpicture}
	\label{fig:ee_encoding}
	}}
	\caption{Different E/E architectures}
	\label{fig:ee}
\end{figure}
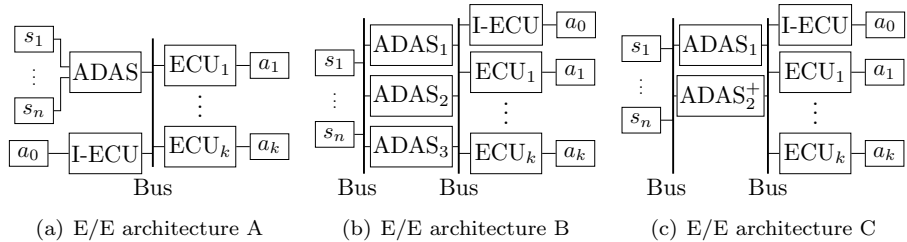

\subsubsection{Partitioning on E/E architecture}
The next design step consists of extending the nominal E/E architecture for vehicle guidance and partitioning the blocks of every safety concept on its elements. The nominal E/E architecture is represented in Fig.~\ref{fig:ee_nominal}. The vehicle guidance function is implemented on an ADAS-platform (Advanced Driver Assistance System) which is connected to all sensors. A number of dedicated ECUs (Electronic Control Unit) control the actuators. On an I-ECU (Integration ECU), additional, non-dedicated actuation functions can be implemented. 
Naturally, implementing all blocks from the safety concepts on the ADAS in Architecture A defeats the purpose of the redundant paths. 

Fig.~\ref{fig:ee} gives further illustrative examples for E/E architectures for the different safety concepts:
For SC1, Architecture B (Fig.~\ref{fig:ee_tmr}) allows an implementation of the three redundant paths on separate ADAS-cores. The Voter could then be implemented on the I-ECU.
For SC2, the following two implementations both yield ASIL D for the safety path, each with TCS and AM on the I-ECU:  
(1) Executing the nominal path on one ADAS and redundant execution of the s-Path on two ADAS-cores in \emph{lock-step mode}, using Architecture B.
(2) \emph{Encoded execution}~\cite{GKSF16} of the s-Path on a single ADAS+-core in Architecture C (Fig.~\ref{fig:ee_encoding}), where the $+$ refers to the additional hardware resources to run an encoded s-Path.
An E/E architecture for SC3 could be realised on Architecture C, where the m-Path is implemented on ADAS$_1$ and the fb-Path on ADAS$_2^{+}$.
Alternatives are considered in our experiments in Sect.~\ref{sec:experiments}.

\subsubsection{Hardware platforms and faults}\label{sec:transientfaultassumption}
We assume that all hardware platforms can completely recover from transient faults (e.g. by restarting the affected path), so that only transient faults directly leading to a system failure are of importance.
Transient faults quickly vanish.
Thus, the probability of an additional transient fault occurring is small.
We assume that this probability is negligible and during a transient fault no other faults occur \cite{iso26262}.
It can be understood as modelling only single transient faults.

\subsection{Measures}\label{Sec:VehicleGuidance_Properties}
The safety goal for the considered systems is to avoid wrong vehicle guidance, i.e., following unsafe trajectories. As the system is designed to be fail-operational, the system should be able to maintain its core functionality for a certain time, e.g. 10 seconds---even in the presence of faults. 
The safety goal is violated, if e.g. two out of three TMR paths or both the n-Path and the s-Path fail. The goal is also violated if e.g. a failure of the n-Path is not detected.
The safety goal is classified as ASIL D. 
We stress that safe faults do not need to be considered. 
For the sake of conciseness, we define the \emph{complement} of probability $p$ as $1-p$.

Several measures allow insights in the safety-performance of the different safety concepts:
\emph{System reliability} refers to the probability that the system safely operates during the considered operational lifetime. To obtain the average failure-probability per hour, the complement of the reliability is scaled with the lifetime.
Besides the reliability, the mean time to failure (MTTF) is a standard measure of interest.
We also consider \emph{degraded states} in which some faults already occurred in the system.
If the system is in a degraded state it still safely operates but provides reduced functionality.
The following measures focusing on degraded states reflect insights also relevant for customer satisfaction:
\begin{compactenum}[1)]
	\item the probability that the system provides the full functionality at time $t$,
	\item the fraction of system failures which occur without being in a degraded state before,
	\item the expected time to failure upon entering a degraded state,
	\item the criticality of a degraded state, in terms of the probability that the system fails within e.g. a typical drive cycle of one hour \cite[5:9.4]{iso26262} while being degraded already, and
	\item the effect on the overall system reliability when imposing limits on the time a system remains operational in a degraded state.
\end{compactenum}
It is important to consider the robustness or \emph{sensitivity} of all measures w.r.t.\ changes in the failure rates.
Furthermore, it is beneficial for a proper analysis to consider the system under (hypothesised) combinations of events.

\section{Technical Background}
\label{sec:technicalbackground}
\subsection{Fault trees}
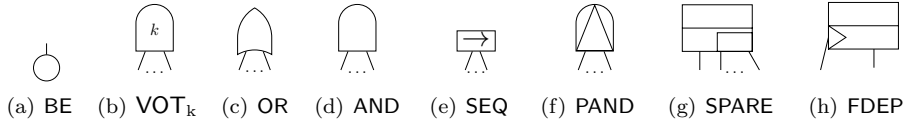
\begin{figure}[tb]
\centering
\subfigure[\BE]{
 \centering
\makebox[0.065\linewidth]{
\scalebox{0.62}{
 \begin{tikzpicture}[  scale=.8,font=\LARGE,text=black, every node/.style={transform shape}, node distance=0.3cm]
	\node[be] (pand) {};
	\node[above=of pand] (output) {};
	\draw[-] (pand) -- (output);
\end{tikzpicture}}}
 \label{Fig:BE} 
}\hfill
\subfigure[\small$\VOT k$]{
 \centering 
\makebox[0.095\linewidth]{
\scalebox{0.62}{
 \begin{tikzpicture}   
    \node[and3] (and) {\rotatebox{270}{$k$}};
    \node[below=0.4 cm of and.input 1, xshift=-0.2cm]  (i1) {};
    \node[below=0.3 cm of and.input 2]  (dots) {$\hdots$};
    
    \node[below=0.4 cm of and.input 3, xshift=0.2cm] (i2) {};
    
    \draw[-] (and.input 1) -- (i1);
    \draw[-] (and.input 3) -- (i2);
  \end{tikzpicture}}}
 \label{Fig:VOT}
}\hfill
   \subfigure[$\OR$]{
  \centering
\scalebox{0.62}{
 \begin{tikzpicture}   
    \node[or3] (and) {};
    \node[below=0.4 cm of and.input 1, xshift=-0.2cm]  (i1) {};
    \node[below=0.3 cm of and.input 2]  (dots) {$\hdots$};
    
    \node[below=0.4 cm of and.input 3, xshift=0.2cm] (i2) {};
    
    \draw[-] (and.input 1) -- (i1);
    \draw[-] (and.input 3) -- (i2);
  \end{tikzpicture}
  }
 \label{Fig:OR} 
}\hfill
 \subfigure[$\AND$]{
  \centering
\makebox[0.085\linewidth]{
\scalebox{0.62}{
 \begin{tikzpicture}   
    \node[and3] (and) {};
    \node[below=0.4 cm of and.input 1, xshift=-0.2cm]  (i1) {};
    \node[below=0.3 cm of and.input 2]  (dots) {$\hdots$};
    
    \node[below=0.4 cm of and.input 3, xshift=0.2cm] (i2) {};
    
    \draw[-] (and.input 1) -- (i1);
    \draw[-] (and.input 3) -- (i2);
  \end{tikzpicture}
  }
  }
 \label{Fig:AND} 
}\hfill
\subfigure[\SEQ]{
  \centering
  \makebox[0.083\linewidth]{
  \scalebox{0.62}{
    \begin{tikzpicture}   
    \node[seq] (and) {$\rightarrow$};
    \node[below=0.4 cm of and.250, xshift=-0.2cm]  (i1) {};
    \node[below=0.3 cm of and.270]  (dots) {$\hdots$};
    
    \node[below=0.4 cm of and.290, xshift=0.2cm] (i2) {};
    
    \draw[-] (and.250) -- (i1);
    \draw[-] (and.290) -- (i2);
  \end{tikzpicture}}}
  \label{Fig:DFTElements_SEQ}
 }\hfill
 \subfigure[\PAND]{
 \centering
\makebox[0.1\linewidth]{
 \scalebox{0.62}{
   \begin{tikzpicture}   
    \node[and3] (and) {};
    \node[triangle,scale=1.62,yshift=-3.5,xscale=0.80] (triangle_a) at (and) {};
    \node[below=0.4 cm of and.input 1, xshift=-0.2cm]  (i1) {};
    \node[below=0.3 cm of and.input 2]  (dots) {$\hdots$};
    
    \node[below=0.4 cm of and.input 3, xshift=0.2cm] (i2) {};
    
    \draw[-] (and.input 1) -- (i1);
    \draw[-] (and.input 3) -- (i2);
  \end{tikzpicture}}}
  \label{Fig:DFTElements_PAND}
  
 }\hfill
 \subfigure[\SPARE]{
\centering
\makebox[0.115\linewidth]{
\scalebox{0.62}{
  \begin{tikzpicture}   
    \node[spare] (and) {};
    \node[below=0.4 cm of and.P]  (i1) {};

    \node[below=0.4 cm of and.SA] (i2) {};
    \node[below=0.3 cm of and.SC] (i3) {$\hdots$};
    
    \node[below=0.4 cm of and.SE, xshift=0.3cm] (i4) {};
    
    \draw[-] (and.P) -- (i1);
    \draw[-] (and.SA) -- (i2);
    \draw[-] (and.SE) -- (i4);
  \end{tikzpicture} }}
  \label{Fig:DFTElements_SPARE}
 }\hfill
 \subfigure[\FDEP]{
   \centering
   \scalebox{0.62}{
      \begin{tikzpicture}   
    
    \node[fdep] (and) {};
    \node[above=0.07cm of and.center] (x) {};
    \node[below=0.7 cm of and.T, xshift=-0.2cm]  (i1) {};
    
    \node[below=0.4 cm of and.EB] (i2) {};
    \draw[-] (and.T) -- (i1);
    \draw[-] (and.EB) -- (i2);
    
  \end{tikzpicture}
  }
    \label{Fig:DFTElements_FDEP}
 }
 \caption{Node types in ((a)-(d)) static and (all) dynamic fault trees}
 \label{Fig:DFTElements}
\end{figure}
Fault trees~\cite{handbook2002,RS15} (FTs) are directed acyclic graphs (DAG) with typed nodes (\AND, \OR, etc.). Nodes of type $T$ are referred to as ``a $T$''. 
Nodes without children (successors in the DAG), are \emph{basic events} (\BE{}s, Fig.\,\ref{Fig:DFTElements}(a)).
Each \BE{} is equipped with some failure rate, or is assumed to not fail by itself (a dummy event).
The \BE{}s of a fault tree $F$ are denoted $F_\BE$. 
Other nodes are \emph{gates} (Fig.\,\ref{Fig:DFTElements}(b)-(h)).
We say a \BE \emph{fails} if the event occurs; a gate ``fails'' if its \emph{failure condition} over its children holds. 
The \emph{top-level event} ($\toplevel{F}$) is a specifically marked node of a FT $F$. $\toplevel{F}$ fails iff the FT $F$ fails.

\subsubsection{Static fault trees}
The key gate for static fault trees (SFTs, gates (b)-(d)) is the \emph{voting} gate (denoted \VOT{$k$}) with \emph{threshold} $k$ and at least $k$ children. 
A \VOT{$k$}-gate fails, if $k$ of its children have failed. 
A \VOT{$1$}-gate equals an \OR-gate, while a \VOT{$k$}-gate with $k$ children equals an \AND-gate.

\subsubsection{Dynamic fault trees}
For fail-operational systems such as future vehicle guidance systems, essential concepts such as (cold) redundancies and complex dependencies cannot be modelled faithfully with static fault trees, cf.\ e.g.~\cite{JGKS16}. 
Dynamic fault trees (DFTs)~\cite{Dugan1990} are an extension well-suited to model and analyse the advanced concepts.
A recent account of precise DFT semantics, including corner cases omitted here, is given in~\cite{JKSV18}. The presentation below is a more intuitive summary of these semantics.
DFTs additionally contain the following gates:

\paragraph{Sequence-enforcers}
The sequence enforcer (\SEQ,~Fig.~\ref{Fig:DFTElements_SEQ}) do not propagate failures, but  restrict the order in which \BE{s} can fail---their children may only fail left-to-right. 
Thus, \SEQ{s} exclude certain failure orders in the model.
Contrary to a widespread belief, \SEQ{s} can in a general context not be modelled by \SPARE{s} (introduced below) \cite{JGKS16}.
For example, \SEQ{s} with children being gates cannot be modelled with \SPARE{s}.
\SEQ{s} appear in the ISO 26262~\cite[10-B.3]{iso26262}, where they are indicated by the boxed L.

\paragraph{Priority-and} The priority-and (\PAND,~Fig.~\ref{Fig:DFTElements_PAND}) fails iff all children fail ordered from left-to-right. 
If the children fail in any other order, e.g.~a child failed before its left sibling, the PAND cannot fail. 

\paragraph{Spare-gates}
Spare-gates (\SPARE{},~Fig.~\ref{Fig:DFTElements_SPARE}) model spare-management and support warm and cold standby. Warm (cold) standby corresponds to a reduced (zero) failure rate.
Likewise to an \AND{}, a \SPARE{} fails if all children have failed. Additionally, the \SPARE \emph{activates} its children from left to right: A child is activated as soon as all children to its left have failed.
By activating and therefore using a child the failure rate is increased.
The children of the \SPARE{s} are assumed to be roots of independent subtrees, these subtrees are called \emph{modules}. Upon activation of the root of a module, the full module is activated.

\paragraph{Functional dependencies}
Functional dependencies (\FDEP,~Fig.~\ref{Fig:DFTElements_FDEP}) ease modelling of feedback-loops. \FDEP{s} have a \emph{trigger} (a node) and a \emph{dependent event} (a \BE). Instead of propagating failure upwards, upon failure of the trigger, the dependent event fails.
While \FDEP{s} are syntactic sugar in SFTs, they cannot be expressed by other gates in DFTs \cite{JGKRS17}.

\paragraph{Activation dependencies}
To overcome syntactic restrictions induced by  \SPARE{s} and to allow greater flexibility with activation, we use \emph{activation dependencies} (\ADEP{s}), as proposed in \cite[Sect.\ 3E]{JGKS16}.
 If the \emph{activation source} is activated, the \emph{activation destination} is also activated.
 We typically use \ADEP{s} in conjunction with an \FDEP{}, where the activation sources are the dependent events and the activation target is the trigger.

\subsection{Markov Chains}
The semantics of DFTs can be expressed in terms of Markov models, more explicitly \emph{continuous-time Markov chains (CTMCs)}~\cite{BHHK10}.

\begin{definition}[CTMC]
A CTMC is a tuple $\ctmc = (\states, \probability, \rate, \atomiclabel)$ with
\begin{compactitem}
    \item $\states$ a finite set of states,
    \item $\probability: \states \times \states \to [0, 1]$ a stochastic matrix with $\sum_{s \in \states'} \probability(s, s')=1$ for all $s \in \states$,
    \item $\rate \colon \states \rightarrow \RRplus$ a function assigning an \emph{exit rate} $\rate(s)$ to each state $s \in \states$,
    \item $\atomiclabel: \states \to 2^{AP}$ a labeling function assigning a set of \emph{atomic propositions} $\atomiclabel(s)$ to each state $s \in \states$.
\end{compactitem}
\end{definition}
The exit rate specifies the rate of a negative exponential distribution governing the residence time in each state.
The \emph{transition rate} between states $s$ and $s'$ is defined as $\rate(s, s')= \rate(s) \cdot \probability(s, s')$.
\emph{State labels} are used in expressing desired properties over CTMCs, and are used here to identify failed or degraded states of the DFT.

\begin{wrapfigure}[7]{r}{0.25\textwidth}
\scalebox{0.7}{
\centering
 \begin{tikzpicture}
  \node[state,initial, initial text=] (s0) [label=225:$\emptyset$] {$s_0$};
  \node[state,below=0.8cm of s0]      (s1) [label=left:$\{a\}$]     {$s_1$};
  \node[state,right=0.8cm of s0]      (s2) [label=right:$\{a\}$]     {$s_2$};
  \node[state,right=0.8cm of s1]      (s3) [label=right:$\{b\}$]     {$s_3$};

  \draw[->] (s0) -- node[right] {3} (s1);
  \draw[->] (s1) -- node[above] {6} (s3);
  \draw[->] (s0) -- node[above] {5} (s2);
  \draw[->] (s2) -- node[right] {4} (s3);
  \draw[->] (s3) -- node[above right] {7} (s0);
 \end{tikzpicture}
 }
 \caption{A CTMC.}
 \label{Fig:ExampleCTMC}
\end{wrapfigure}

\paragraph{Example}
In Fig.~\ref{Fig:ExampleCTMC} a CTMC with 4 states and the transition rates is given.
The corresponding exit rates and transition probabilities can be derived from the transition rates.
The exit rate for $s_0$ is $3 + 5 = 8$.
The transition probabilities are $\probability(s_0, s_1)=\nicefrac{3}{8}$ and $\probability(s_0, s_2)=\nicefrac{5}{8}$.
State $s_2$ is labelled with $a$, state $s_3$ is labelled with $b$.

\section{Creating the Dynamic Fault Trees}
\label{sec:modelling}
This section describes in detail the approach from Fig.~\ref{fig:overview}. 
In particular, we discuss how to systematically create an FT consisting of three layers.
The top-level event is assumed to represent a safety-critical failure of the vehicle-guidance system. 
From top to bottom, the layers are:
\begin{compactenum}
	\item the \emph{system layer} describes how the top-level event occurs due to failing function blocks (e.g., EP or TP). Thus, the layer's basic events describe the failure of a function block.
	\item the \emph{block layer} describes how a function block can fail, taking into account the possibility of failing computation, or wrong inputs. Thus, the layer's basic events describe either failures of the E/E-architecture. In particular, either failures of the hardware platform on which the function block is executed, or the failure of busses that realise the communication between function blocks.
	\item the \emph{hardware layer} describes how the hardware platforms can fail, based on failures of hardware components, which are considered as basic events.
\end{compactenum}
The system- and block layer depend on the functional architecture of the system, in particular on the structure and information captured in a (given) functional block diagram.
The fault trees for the hardware are typically constructed by the manufacturers.
The connections between the block layer and the hardware layer are inferred from the E/E-architecture and the hardware partitioning.

\subsection{From Function Block Diagram to the system- and block layer}
\label{Subsec:FBD2FFT}

We discuss the creation of the system- and block layer of the fault tree, based on a functional architecture and a function block diagram for this architecture.
\begin{definition}[Block diagram]
A \emph{block diagram} $\blockdiagram = (\blocks, \channels)$ is a finite directed graph. The vertices $\blocks$ are a set of \emph{blocks}, the edges $\channels$ are called \emph{channels}. 
\end{definition}
Given a channel $(B, B') \in \channels$, we call $B$ the source and $B'$ the target. 
The set $\inchannels{B} = \{ e \in \channels \mid e \in \blocks \times \{ B \} \}$ denotes the input channels of block $B$, and $\outchannels{B} = \{ e \in \channels \mid e \in \{B\} \times \blocks \}$ denotes its output channels. 
The block diagram may contain cycles, cf.~Fig.~\ref{fig:feedbackloop}.

Within the scope of this paper, we advocate the structured manual creation of the system- and block layer. 
A further discussion is given in Sect.~\ref{sec:manual_creation}.
Below, we discuss the creation of both layers.

\begin{figure}
\centering
\begin{tikzpicture}
\node [rectangle, draw] (x1) at (1.4, 5.2) {$B_1$};
	\node [rectangle, draw] (x2) at (1.4, 4.6) {$B_2$};
	\node [rectangle, draw] (x3) at (5.4, 4.6) {$B_3$};
	\draw[->] (x1) -| (x3);
	\draw[->] (x2) -- (x3);
\end{tikzpicture}
	\caption{Toy-example function block diagram}
	\label{fig:toyexample_fbd}
\end{figure}
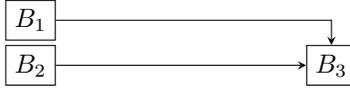

\begin{example}
Consider the function block diagram in Fig.~\ref{fig:toyexample_fbd}. 
Formally, the diagram is given by 
$\blocks = \{ B_1, B_2, B_3 \}$, and 
$\channels = \{ (B_1, B_3), (B_2, B_3) \}$.
\end{example}

\paragraph{Remark}
Typically, fault trees are created top-down. 
We  deviate from this scheme to ease the presentation, and show the potential for automation.

\subsubsection{Creation of the block layer}
For the systematic construction of the block layer, we assume that the possible causes of a block to fail are described by an appropriate block fault tree. 
The causes include faulty input channels.

\paragraph{Creating fault trees for each block}

In a first step, we create a fault tree $F^B$ for each block $B \in \blocks$ in the block diagram.
To reflect the interaction within both the design pattern and the influence from the assigned hardware later on, we annotate each block fault tree $F^B$ with relevant information.
\begin{compactitem}
\item For each input channel of $B$, we define a \BE for faulty input.
\item For each output channel $c$ of $B$, we define an element in $F^B$ which causes $B$'s output on $c$ to be faulty. Later, we propagate the faulty output to the block fault trees for the target block along the channel $c$.
\item We define a \BE for the failure of the hardware platform that executes $B$. 
\end{compactitem}
Formally, we capture these fault trees as follows.
\begin{definition}[Block fault tree]
A block fault tree for block $B$ is a tuple $(F^B, I^B, O^B, Y^B)$, where $F^B$ is a fault tree, $I^B\colon \inchannels{B} \to F^B_\BE$ denotes \emph{input faults}, $O^B\colon \outchannels{B} \to F^B$ \emph{output failures}, and $Y^B \in F^B_\BE$ the \emph{hardware fault} \BE.
\end{definition}

\begin{figure}[t]
\centering
\subfigure[Block FT]{
\scalebox{0.750}{
	\begin{tikzpicture}[scale=.4,text=black]
\centering
    \node[or3] (and) {};
    
	\node[labelbox] (and_label) at (and.east) {\underline{$\text{block}$}};

	\node[or2,below=1.5cm of and] (or2) {};
	\node[labelbox] (in_label) at (or2.east) {input};

	\node[be,below=1.3cm of and.center,xshift=-0.1cm] (B) {};
	\node[labelbox] (b_label) at (B.north) {intern};
	\node[be,left=0.2cm of B] (hw) {};
	\node[labelbox] (hw_label) at (hw.north) {hw};
	
	\node[be, below=3.0cm of and.center] (D) {};
	\node[labelbox] (d_label) at (D.north) {in 1};
	\node[be, below=3.0cm of and.center, xshift=0.8cm] (C) {};
	\node[labelbox] (c_label) at (C.north) {in 2};

	\draw[-] (and.input 1) -- (hw_label.north);
	\draw[-] (and.input 2) -- (b_label.north);
	\draw[-] (and.input 3) -- (in_label.north);
	\draw[-] (or2.input 1) -- (d_label.north);
	\draw[-] (or2.input 2) -- (c_label.north);
	
\end{tikzpicture}}
\label{fig:minft}
}\qquad\qquad
\subfigure[Block FT -- Voter]{
\scalebox{0.750}{
	\begin{tikzpicture}[scale=.4,text=black]
\centering
    \node[or3] (and) {};
    
	\node[labelbox] (and_label) at (and.east) {\underline{$\text{block}$}};

	\node[and3,below=1.5cm of and] (or2) {\rotatebox{270}{$2$}};
	\node[labelbox] (in_label) at (or2.east) {input};

	\node[be,below=1.3cm of and.center,xshift=-0.1cm] (B) {};
	\node[labelbox] (b_label) at (B.north) {intern};
	\node[be,left=0.2cm of B] (hw) {};
	\node[labelbox] (hw_label) at (hw.north) {hw};
	
	\node[be, below=3.0cm of and.center] (D) {};
	\node[labelbox] (d_label) at (D.north) {in 1};
	\node[be, below=3.0cm of and.center, xshift=0.8cm] (C) {};
	\node[labelbox] (c_label) at (C.north) {in 2};
	\node[be, below=3.0cm of and.center, xshift=1.8cm] (E) {};
	\node[labelbox] (e_label) at (E.north) {in 3};

	\draw[-] (and.input 1) -- (hw_label.north);
	\draw[-] (and.input 2) -- (b_label.north);
	\draw[-] (and.input 3) -- (in_label.north);
	\draw[-] (or2.input 1) -- (d_label.north);
	\draw[-] (or2.input 2) -- (c_label.north);
	
	\draw[-] (or2.input 3) -- (e_label.north);
\end{tikzpicture}}
\label{fig:ft_voter}
}\qquad\qquad
\subfigure[Block FT -- Switch]{
\scalebox{0.750}{
	\begin{tikzpicture}[scale=.4,text=black]
\centering
    \node[or2] (and) {};
    
	\node[labelbox] (and_label) at (and.east) {\underline{$\text{switch}$}};

	\node[and2,below=1.7cm of and, yshift=-0.2cm] (or2) {};
	\node[labelbox] (in_label) at (or2.east) {all input wrong};
	
	\node[and2,below=1.7cm of and, yshift=1.8cm] (and3) {};
	\node[labelbox] (sw_label) at (and3.east) {wrong path};\node[triangle,scale=1.62,yshift=-3.5,xscale=0.80] (triangle_a) at (and3) {};
	
	\node[be,below=3.0cm of and.center] (D) {};
	\node[labelbox] (d_label) at (D.north) {in 1};
	\node[be,left=0.4cm of D] (B) {};
	\node[labelbox] (b_label) at (B.north) {switching};
	\node[be,right=0.4cm of D] (C) {};
	\node[labelbox] (c_label) at (C.north) {in 2};
	
	\draw[-] (and.input 1) -- (sw_label.north);
	\draw[-] (and.input 2) -- (in_label.north);
	
	\draw[-] (or2.input 1) -- (d_label.north);
	\draw[-] (or2.input 2) -- (c_label.north);
	\draw[-] (and3.input 1) -- (b_label.north);
	\draw[-] (and3.input 2) -- (d_label.north);
\end{tikzpicture}}%
\label{fig:switching}%
}
\caption{Block fault trees}
\end{figure}
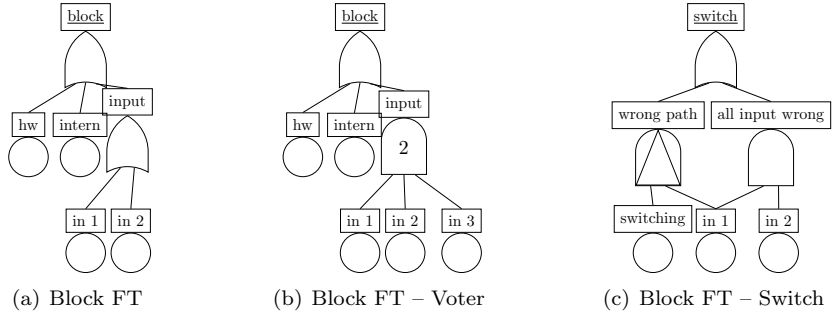

A fault tree for a block $B$ with two input channels ($c_1,c_2$) and one output channel ($c_3$) is given in Fig.~\ref{fig:minft}.
The block fails if either the hardware fails, due to some internal fault, or due to faulty input. 
Typically, the hardware failure (formally, $Y^B = \textsf{hw}$) is connected to the hardware fault tree, see Sect.~\ref{Subsec:FT_hardware}.
The internal fault can be used to support sub-blocks or for additional fallible events.
Faulty input means the failure of any of the inputs (formally, $I^B(c_1) = \textsf{in 1}$, $I^B(c_2) = \textsf{in 2}$). 
The output is faulty, if the block fails (formally, $O^B(c_3) = \textsf{block}$).
Adaptions of the standard scheme are possible:
\begin{compactitem}
\item 
For the voter block, as in triple modular redundancy SC1. We assume that the voter fails if two out of three inputs are faulty. Thus, we obtain a block fault tree as depicted in Fig.~\ref{fig:ft_voter}, with three inputs, and the block fails if two out of three inputs fail, as reflected by a $\VOT{2}$-gate.
\item 
For the switch as in SC3. The switching mechanism may fail. This only leads to a failure when the path has to be switched. 
This behaviour is reflected by the DFT in Fig.~\ref{fig:switching}: the Switch fails if it either uses the wrong path or if all input is wrong. 
The path is wrong if the switching mechanism fails before the primary input fails, i.e., it can no longer switch to an operational path, as reflected by a \PAND-gate.
\end{compactitem}
Block fault trees can easily be extended to several hardware failures, to input/output faults of several types, etc.

\begin{figure}
\centering
\subfigure[Simple block diagram with connected block fault tree]{
\scalebox{0.75}{
\begin{tikzpicture}
	\node [rectangle, draw] (x1) at (1.4, 5.2) {$B_1$};
	\node [rectangle, draw] (x2) at (1.4, 4.6) {$B_2$};
	\node [rectangle, draw] (x3) at (5.4, 4.6) {$B_3$};
	\draw[->] (x1) -| (x3);
	\draw[->] (x2) -- (x3);

	 \node[or2] (b1) at (1,1.5) {};
	 \node[labelbox] (b1_label) at (b1.east) {$B_1$};
	
    \node[or2] (b2) at (3,1.5) {};
    \node[labelbox] (b2_label) at (b2.east) {$B_2$};

    \node[or2] (and) at (6,1.5) {};
    
	\node[labelbox] (and_label) at (and.east) {$B_3$};

	\node[or2,below=1.5cm of and, yshift=1.6cm] (or2) {};
	\node[labelbox] (in_label) at (or2.east) {input};

    		\node[be,below=1cm of and, xshift=-0.2cm] (hw) {};
	\node[labelbox] (hw_label) at (hw.north) {hw};
	
	\node[be, below=1.4cm of or2.center, xshift=-0.8cm] (D) {};
	\node[labelbox] (d_label) at (D.north) {in 2};
	\node[be, below=1.4cm of or2.center, xshift=0.8cm] (C) {};
	\node[labelbox] (c_label) at (C.north) {in 1};
	
	\node[fdep, scale=0.8] (f1) at (1.7, 3) {};
    \node[fdep, scale=0.8] (f2) at (3.7, 3) {};
    
    \node[be] (1a) at (0.5, 0.1) {};
	\node[labelbox] (1a_label) at (1a.north) {hw};
	\node[be] (1b) at (1.5, 0.1) {};
	\node[labelbox] (1b_label) at (1b.north) {intern};
	\node[be] (2a) at (2.5, 0.1) {};
	\node[labelbox] (2a_label) at (2a.north) {hw};
	\node[be] (2b) at (3.5, 0.1) {};
	\node[labelbox] (2b_label) at (2b.north) {intern};
	
	\draw (b1.input 1) -- (1a_label.north);
	\draw (b1.input 2) -- (1b_label.north);
	\draw (b2.input 1) -- (2a_label.north);
	\draw (b2.input 2) -- (2b_label.north);
	
	\draw[dashed] (0.5, 3.7) -- (6.3, 3.7);
	
	\draw (f1.T) -- (b1_label.north);
	\draw (f1.ED) -- (c_label.north);
	\draw (f2.T) -- (b2_label.north);
	\draw (f2.ED) -- (d_label.north);

	\draw[-] (and.input 1) -- (in_label.north);
	\draw[-] (and.input 2) -- (hw_label.north);
		\draw[-] (or2.input 1) -- (d_label.north);
	\draw[-] (or2.input 2) -- (c_label.north);

\end{tikzpicture}}
\label{fig:connectedblocksft}
}\qquad\qquad
\subfigure[Feedback loop and connected block FT]{
\scalebox{0.75}{
\begin{tikzpicture}
	\node [rectangle, draw] (x1) at (1.4, 4.6) {$B_1$};
	\node [rectangle, draw] (x2) at (3.4, 4.6) {$B_2$};
	\node [rectangle, draw] (x3) at (5.4, 4.6) {$B_3$};
	\draw[->] (x1) -- (x2);
	\draw[->] (x2) -- (x3);
	\draw[->] (x3.south) -- +(0, -0.3) -| (x1.south);

	 \node[or2] (b1) at (1,1.5) {};
	 \node[labelbox] (b1_label) at (b1.east) {$B_1$};
	
    \node[or2] (b2) at (3,1.5) {};
    \node[labelbox] (b2_label) at (b2.east) {$B_2$};
	
    \node[or2] (b3) at (5,1.5) {};
    \node[labelbox] (b3_label) at (b3.east) {$B_3$};
	
	\node[fdep, scale=0.8] (f1) at (1.7, 3) {};
    \node[fdep, scale=0.8] (f2) at (3.7, 3) {};
    \node[fdep, scale=0.8] (f3) at (5.7, 3) {};
    
    \node[be] (1a) at (0.4, 0.1) {};
	\node[labelbox] (1a_label) at (1a.north) {in 3};
	\node[be] (1b) at (1.6, 0.1) {};
	\node[labelbox] (1b_label) at (1b.north) {hw};
	\node[be] (2a) at (2.4, 0.1) {};
	\node[labelbox] (2a_label) at (2a.north) {in 1};
	\node[be] (2b) at (3.6, 0.1) {};
	\node[labelbox] (2b_label) at (2b.north) {hw};
	\node[be] (3a) at (4.4, 0.1) {};
	\node[labelbox] (3a_label) at (3a.north) {in 2};
	\node[be] (3b) at (5.6, 0.1) {};
	\node[labelbox] (3b_label) at (3b.north) {hw};

	\phantom{\node [rectangle, draw] (ph1) at (1.4, 5.2) {};
	\node[be, below=1.2cm of 2b.center, xshift=0.8cm] (ph2) {};}
	
	\draw (b1.input 1) -- (1a_label.north);
	\draw (b1.input 2) -- (1b_label.north);
	\draw (b2.input 1) -- (2a_label.north);
	\draw (b2.input 2) -- (2b_label.north);
	\draw (b3.input 1) -- (3a_label.north);
	\draw (b3.input 2) -- (3b_label.north);
	
	\draw[dashed] (0.5, 3.7) -- (6.3, 3.7);
	
	\draw (f1.T) -- (b1_label.north);
	\draw (f1.ED) -- (2a_label.north);
	\draw (f2.T) -- (b2_label.north);
	\draw (f2.ED) -- (3a_label.north);
	\draw (f3.T) -- (b3_label.north);
	\draw (f3.ED) -- (5.7, 1) -- (1a_label.north);

\end{tikzpicture}}
\label{fig:feedbackloop}
}
\caption{Connecting block fault trees}
\end{figure}
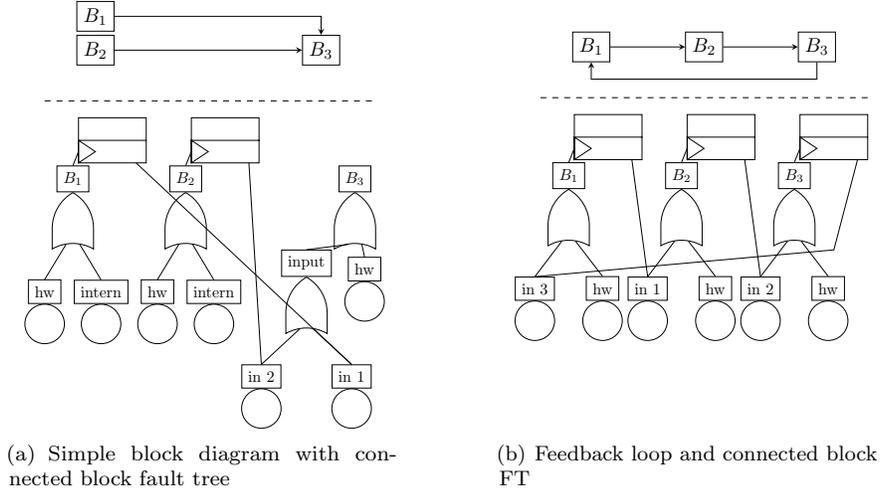
\paragraph{Connecting block fault trees}
The goal of this step is to connect the inputs and outputs of the block fault trees as specified by the relation $\channels$.
The \emph{connected block fault tree} consists of the disjoint union of all block fault trees. 
Additionally, for each channel $c = (B,B')$ in the block diagram, we connect the output failure $O^B(c)$ with the input fault $I^{B'}(c)$.
The connection is realised by means of an \FDEP with trigger $O^B(c)$ and dependent event $I^{B'}(c)$.
As we only consider the block layer, we do not have a top-level event.

\begin{example}\label{ex:connectedft}
We illustrate the connected block fault tree in Fig.~\ref{fig:connectedblocksft} for a toy example.
For each block, we use a standard block as in Fig.~\ref{fig:minft}.
We assume faulty input to the incoming channels of $B_3$ if the respective source block  fails.
\end{example}

A major benefit of \FDEP{}s is that they allow to faithfully model feedback loops~\cite{handbook2002}. 
We illustrate the support for feedback loops in Fig.~\ref{fig:feedbackloop}: 
For three blocks as shown on top, the three DFTs are connected via \FDEP{s}. 
If, e.g., $B_1$ fails, the failure is propagated to the input of $B_2$, etc. 
Using \FDEP{s} allows that cyclic dependencies can be modelled by (acyclic) fault trees, and is very flexible.

\subsubsection{Creation of the system layer}

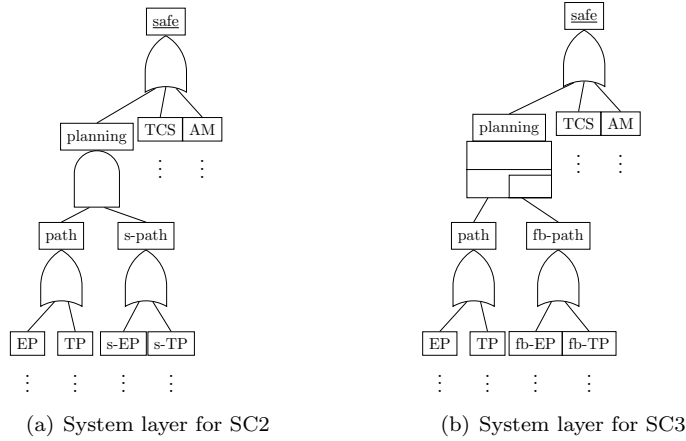
\begin{figure}\centering
\subfigure[System layer for SC2]{
\scalebox{0.750}{
	\begin{tikzpicture}[scale=.4,text=black]
\centering
    \node[or3] (and) {};
    \node[left=2cm of and] () {};
    
    \node[right=2cm of and] () {};
	\node[labelbox] (and_label) at (and.east) {\underline{$\text{safe}$}};

	\node[and2,below=2cm of and, yshift=2cm] (or2) {};
	\node[labelbox] (in_label) at (or2.east) {planning};

	\node[below=1.3cm of and.center,xshift=-0.1cm] (B) {$\vdots$};
	\node[labelbox] (b_label) at (B.north) {TCS};
	\node[right=0.4cm of B] (hw) {$\vdots$};
	\node[labelbox] (hw_label) at (hw.north) {AM};
	
	\node[or2, below=1.8cm of or2.center, yshift=1cm] (D) {};
	\node[labelbox] (d_label) at (D.east) {path};
	\node[or2, below=1.8cm of or2.center, yshift=-0.5cm] (C) {};
	\node[labelbox] (c_label) at (C.east) {s-path};
	
	\node[below=1.3cm of D.center,xshift=-0.6cm] (B2) {$\vdots$};
	\node[labelbox] (b2_label) at (B2.north) {EP};
	\node[right=0.5cm of B2] (hw2) {$\vdots$};
	\node[labelbox] (hw2_label) at (hw2.north) {TP};
	
	\node[below=1.3cm of C.center,xshift=-0.4cm] (B3) {$\vdots$};
	\node[labelbox] (b3_label) at (B3.north) {s-EP};
	\node[right=0.5cm of B3] (hw3) {$\vdots$};
	\node[labelbox] (hw3_label) at (hw3.north) {s-TP};
		
	\draw[-] (and.input 3) -- (hw_label.north);
	\draw[-] (and.input 2) -- (b_label.north);
	\draw[-] (and.input 1) -- (in_label.north);
	\draw[-] (or2.input 1) -- (d_label.north);
	\draw[-] (or2.input 2) -- (c_label.north);
	
	\draw[-] (D.input 1) -- (b2_label.north);
	\draw[-] (D.input 2) -- (hw2_label.north);
	\draw[-] (C.input 1) -- (b3_label.north);
	\draw[-] (C.input 2) -- (hw3_label.north);

\end{tikzpicture}}
\label{fig:syssc2}
}\qquad\qquad
\subfigure[System layer for SC3]{
\scalebox{0.750}{
	\begin{tikzpicture}[scale=.4,text=black]
\centering
    \node[or3] (and) {};
    \node[left=2cm of and] () {};
    
    \node[right=2cm of and] () {};
	\node[labelbox] (and_label) at (and.east) {\underline{$\text{safe}$}};

	\node[spare,below=1.4cm of and, xshift=-1.7cm] (or2) {};
	\node[labelbox] (in_label) at (or2.north) {planning};

	\node[below=1.3cm of and.center,xshift=-0.1cm] (B) {$\vdots$};
	\node[labelbox] (b_label) at (B.north) {TCS};
	\node[right=0.4cm of B] (hw) {$\vdots$};
	\node[labelbox] (hw_label) at (hw.north) {AM};
	
	\node[or2, below=2cm of or2.center, yshift=1cm] (D) {};
	\node[labelbox] (d_label) at (D.east) {path};
	\node[or2, below=2cm of or2.center, yshift=-0.5cm] (C) {};
	\node[labelbox] (c_label) at (C.east) {fb-path};
	
	\node[below=1.3cm of D.center,xshift=-0.6cm] (B2) {$\vdots$};
	\node[labelbox] (b2_label) at (B2.north) {EP};
	\node[right=0.5cm of B2] (hw2) {$\vdots$};
	\node[labelbox] (hw2_label) at (hw2.north) {TP};
	
	\node[below=1.3cm of C.center,xshift=-0.4cm] (B3) {$\vdots$};
	\node[labelbox] (b3_label) at (B3.north) {fb-EP};
	\node[right=0.6cm of B3] (hw3) {$\vdots$};
	\node[labelbox] (hw3_label) at (hw3.north) {fb-TP};

	\draw[-] (and.input 3) -- (hw_label.north);
	\draw[-] (and.input 2) -- (b_label.north);
	\draw[-] (and.input 1) -- (in_label.north);
	\draw[-] (or2.P) -- (d_label.north);
	\draw[-] (or2.SB) -- (c_label.north);
	
		\draw[-] (D.input 1) -- (b2_label.north);
	\draw[-] (D.input 2) -- (hw2_label.north);
	\draw[-] (C.input 1) -- (b3_label.north);
	\draw[-] (C.input 2) -- (hw3_label.north);

\end{tikzpicture}}
\label{fig:syssc3}
}
\caption{Task-based construction of the system layer}
\end{figure}

The goal of this step is to express the occurrence of a safety-critical failure as a fault tree over the failure of the blocks.
Towards this goal, we process top-down.
A \emph{task-based} partitioning of the DFT is helpful.
 The  fault tree, i.e.\ the top-level event, is assumed to fail if any of the tasks can no longer be executed (\OR). 
The tasks fail if no block can realise the task anymore (\AND). 
We do not need to consider error propagation through the blocks at this point, as this is already handled by the  connections in the connected blocks fault tree.
We consider the system layer for SC2 in Fig.~\ref{fig:syssc2}: The system fails, if either the planning, the selection, or the actuator management fails. Planning fails if both the nominal and the safety path fail. 

To model cold/warm redundancy, we replace \AND by the more general \SPARE-gate.
Let us illustrate this for SC3. Recall that the fb-Path is in cold standby. 
The traction and environment in the fallback-path only operate as soon as they are required, i.e., as soon as the nominal path fails. This behaviour is captured by using a \SPARE instead of an \AND, as in Fig.~\ref{fig:syssc3}. 

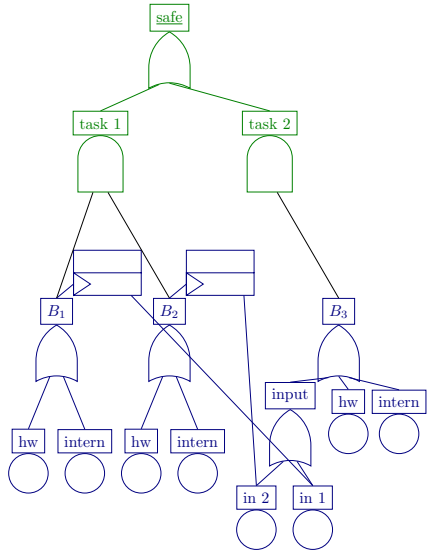
\begin{figure}
\centering
	\scalebox{0.750}{
	\begin{tikzpicture}
        \colorlet{blo}{darkblue}
        \colorlet{sys}{darkgreen}
        \colorlet{connec}{black}

    \node[or3,sys] (top) at (3,7) {};
	\node[labelbox,sys] (top_label) at (top.east) {\underline{$\text{safe}$}};
	\node[and2,sys,below=1.8cm of top, yshift=2cm] (top2) {};
	\node[labelbox,sys] (top2_label) at (top2.east) {task 1};

	\node[and2,sys,below=1.8cm of top, yshift=-1cm] (top3) {};
	\node[labelbox,sys] (top3_label) at (top3.east) {task 2};

	\node[or2,blo] (b1) at (1,1.8) {};
	 \node[labelbox,blo] (b1_label) at (b1.east) {$B_1$};
	
    \node[or2,blo] (b2) at (3,1.8) {};
    \node[labelbox,blo] (b2_label) at (b2.east) {$B_2$};

    \node[or3,blo] (and) at (6,1.8) {};
	\node[labelbox,blo] (b3_label) at (and.east) {$B_3$};

	\node[or2,blo,below=1.5cm of and, yshift=1.6cm] (or2) {};
	\node[labelbox,blo] (in_label) at (or2.east) {input};
	
    \node[be,draw=blo,below=1cm of and, xshift=0.7cm] (intern) {};
	\node[labelbox,blo] (intern_label) at (intern.north) {intern};
	\node[be,draw=blo,below=1cm of and, xshift=-0.2cm] (hw) {};
	\node[labelbox,blo] (hw_label) at (hw.north) {hw};
	
	\node[be,draw=blo, below=1.2cm of or2.center, xshift=-0.6cm] (D) {};
	\node[labelbox,blo] (d_label) at (D.north) {in 2};
	\node[be,draw=blo, below=1.2cm of or2.center, xshift=0.4cm] (C) {};
	\node[labelbox,blo] (c_label) at (C.north) {in 1};
	
	\node[fdep,blo, scale=0.8] (f1) at (1.9, 3.3) {};
    \node[fdep,blo, scale=0.8] (f2) at (3.9, 3.3) {};
    
    \node[be,draw=blo] (1a) at (0.5, 0.1) {};
	\node[labelbox,blo] (1a_label) at (1a.north) {hw};
	\node[be,draw=blo] (1b) at (1.5, 0.1) {};
	\node[labelbox,blo] (1b_label) at (1b.north) {intern};
	\node[be,draw=blo] (2a) at (2.5, 0.1) {};
	\node[labelbox,blo] (2a_label) at (2a.north) {hw};
	\node[be,draw=blo] (2b) at (3.5, 0.1) {};
	\node[labelbox,blo] (2b_label) at (2b.north) {intern};
	
	\draw[-,blo] (b1.input 1) -- (1a_label.north);
	\draw[-,blo](b1.input 2) -- (1b_label.north);
	\draw[-,blo](b2.input 1) -- (2a_label.north);
	\draw[-,blo](b2.input 2) -- (2b_label.north);

	\draw[-,blo](f1.T) -- (b1_label.north);
	\draw[-,blo](f1.ED) -- (c_label.north);
	\draw[-,blo](f2.T) -- (b2_label.north);
	\draw[-,blo](f2.ED) -- (d_label.north);

	\draw[-,blo] (and.input 1) -- (in_label.north);
	\draw[-,blo] (and.input 2) -- (hw_label.north);
	\draw[-,blo] (and.input 3) -- (intern_label.north);
	\draw[-,blo] (or2.input 1) -- (d_label.north);
	\draw[-,blo] (or2.input 2) -- (c_label.north);
	\draw[-,sys] (top.input 1) -- (top2_label.north);
	\draw[-,sys] (top.input 2) -- (top3_label.north);
	\draw[-] (top3.input 2) -- (b3_label.north);
	\draw[-] (top2.input 1) -- (b1_label.north);
	\draw[-] (top2.input 2) -- (b2_label.north);
	
\end{tikzpicture}}
\caption{Putting the system layer (green) and the block layer (blue) together}
\label{fig:ex_sys_and_blocks}
\end{figure}

\begin{example}\label{ex:functionalft}
Let us continue Example~\ref{ex:connectedft}. 
Assume that the blocks $B_1$ and $B_2$ execute the same task with $B_2$ in hot standby, and $B_3$ executes another task. 
The resulting DFT is depicted in Fig.~\ref{fig:ex_sys_and_blocks}.
Green colour indicates the system layer and blue colour the block layer.
\end{example}

\subsection{Adding the hardware layer to the system- and block layers}
In this section, we discuss how to combine the system- and block layer fault tree with the hardware layer.
We assume additional input fault trees for all hardware components, an E/E-architecture and a hardware assignment. 
Below, we first discuss the  additional inputs and then consider how to combine them to a complete fault tree.

\subsubsection{Fault trees for hardware platforms and buses}
\label{Subsec:FT_hardware}
We assume that the (D)FTs to model hardware failures are provided by the manufacturers, and we do not make any assumptions about the structure of these FTs.  
We briefly illustrate how to integrate coverage and both transient and permanent faults in DFTs, analogous to~\cite[10:B]{iso26262}.
We would like to stress the dynamic nature of coverage by fallible safety mechanisms.
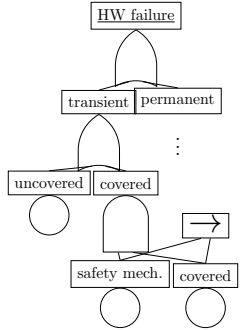
\begin{figure}
\centering
\scalebox{0.75}{
	\begin{tikzpicture}[scale=.4,text=black]
\centering
    \node[or2] (top) {};
    
	\node[labelbox] (top_label) at (top.east) {\underline{$\text{HW failure}$}};

	\node[color=white,or2,below=1.5cm of top] (perm) {};
	\node[labelbox] (perm_label) at (perm.east) {permanent};
    \node[below=-0.5 cm of perm.center] (dots) {$\vdots$};
	
	\node[or2,below=1.5cm of top, yshift=1.4cm] (trans) {};
	\node[labelbox] (trans_label) at (trans.east) {transient};
	
	\node[and2,below=1.4cm of trans, yshift=0.3cm] (and) {};
	\node[labelbox] (and_label) at (and.east) {covered};
	\node[be,left=1.0cm of and, yshift=0.66cm] (uncov) {};
	\node[labelbox] (uncov_label) at (uncov.north) {uncovered};
	\node[seq,right=1.0cm of and.center] (seq) {$\rightarrow$};
	
	\node[be,below=1.1cm of and.center, xshift=-0.1cm] (safety) {};
	\node[labelbox] (safety_label) at (safety.north) {safety mech.};
	\node[be,right=0.8cm of safety] (cov) {};
	\node[labelbox] (cov_label) at (cov.north) {covered};
	
	\draw[-] (top.input 1) -- (trans_label.north);
	\draw[-] (top.input 2) -- (perm_label.north);
	
	\draw[-] (trans.input 1) -- (uncov_label.north);
	\draw[-] (trans.input 2) -- (and_label.north);
	\draw[-] (and.input 1) -- (safety_label.north);
	\draw[-] (and.input 2) -- (cov_label.north);
	\draw[-] (seq.250) -- (safety_label.north);
	\draw[-] (seq.290) -- (cov_label.north);
\end{tikzpicture}
}	   
\caption{Hardware Fault Tree}
\label{fig:hw_tree}
\end{figure}
  
Consider the DFT depicted in Fig.~\ref{fig:hw_tree}. 
A failure results from either a transient or permanent error. 
Each type has its own corresponding safety mechanism. 
We consider the transient error. 
A transient error occurs if either the transient fault is not covered by the safety mechanism, or the fault is covered but the safety mechanism has failed before. 
We model the latter by a \SEQ.\footnote{For other analysis purposes a model using a \PAND might be more adequate.}
It can neither be modelled faithfully by a static FT nor a \PAND.
A \PAND would model that a covered error is never propagated if the covered fault occurs before.
The permanent error is modelled similarly.

\subsubsection{Hardware Assignment and E/E Architecture}
\label{Subsec:Hardware_assignment}
We briefly discuss the assumptions about the hardware assignment and the E/E architecture.
We formalise the concepts as follows:

\begin{definition}[E/E-architecture]
	An E/E-architecture is a tuple $(\hardwareplatforms, \buses)$ of a set $\hardwareplatforms$  of hardware platforms, and a set $\buses$ of buses.
	 Each bus is a transitive relation over hardware platforms, i.e. $\buses \subseteq 2^{\hardwareplatforms \times \hardwareplatforms}$.
	  \end{definition}
Often, buses are equivalence relations, i.e., they connect a set of platforms to each other, as e.g. in FlexRay~\cite{isoFlexRay} or CAN~\cite{isoCAN}.	  
We assume that all hardware platforms $p \in \hardwareplatforms$ contain an internal bus $\internal{p}$ = $\{ (p,p) \} \in \buses$. 
Consequently, function blocks on the same hardware platform can always communicate.

\begin{example}\label{ex:eearchitecture}
We formalise the E/E-architecture B from Fig.~\ref{fig:ee_tmr}.	
The hardware platforms $\hardwareplatforms$ are: 
\[ \{ s_1, \hdots, s_n, \text{ADAS}_1, \text{ADAS}_2, \text{ADAS}_3, \text{I-ECU}, \text{ECU}_1, \hdots, \text{ECU}_k, a_0, \hdots, a_k \}, \]
and the $\buses$ is given as 
\[ \{ \text{ECU-actuator}_0, \hdots, \text{ECU-actuator}_k, \text{CAN-BUS}\} \cup \{\internal{p} \mid  p \in \hardwareplatforms \}, \]
with \begin{compactitem}
 \item 
 $\text{ECU-actuator}_0 = \{ (\text{I-ECU}, a_0), (a_0, \text{I-ECU}) \}$, 
 \item 
$\text{ECU-actuator}_i = \{  (\text{ECU}_i, a_i), (a_i, \text{ECU}_i) \}$, and 
\item $\text{CAN-BUS} = H \setminus \{ a_0, \hdots, a_k \} \times H \setminus \{ a_0, \hdots, a_k \}$. 
 \end{compactitem}
\end{example}

\begin{definition}[Hardware assignment]
Given a block diagram $\blockdiagram = (\blocks, \channels)$ and an E/E-architecture $(\hardwareplatforms, \buses)$, a \emph{hardware-assignment} is a function $\hardwareassignment\colon \blockdiagram \rightarrow \hardwareplatforms \cup \buses$ s.t. $\hardwareassignment(\blocks) \subseteq \hardwareplatforms$ and $\hardwareassignment(\channels) \subseteq \buses \cup \hardwareplatforms$.
\end{definition}
More precisely, hardware assignments map internal channels to hardware platforms, and all other channels to buses. 
A hardware assignment is \emph{consistent}, if the source and target of each channel are connected by a bus. Thus, the assignment is consistent if \[ (B,B') \in \channels \implies \exists C \in \buses \text{ with } (h(B), h(B')) \in C. \]

A hardware assignment trivially supports mapping several function blocks to the same hardware.
In this case, we assume that a failure of a function block does not affect other function blocks on the same hardware.
If necessary, dependent failures of function blocks can be modelled explicitly in the DFT using \FDEP{s}.
We require that any function block runs on at most one hardware platform. 
If the same function should be implemented on several hardware platforms, we require an explicit duplication in the function block diagram. 

\paragraph{Remark}
We do not consider whether a hardware assignment is feasible, i.e., whether the hardware platforms provide the (computational) resources to realise the functions.

\paragraph{Block-based assignments}
Typically, we are only given an assignment from blocks to hardware platforms; the channel-assignment then follows from the E/E architecture. 
That is, we assume that any channel between two blocks assigned to the same hardware platform is realised by the internal bus.
For channels that connect blocks which are assigned to different platforms, we select the unique bus (e.g., the CAN-bus) that connects these two platforms.
If there is no unique bus, then manual intervention is required to specify the proper bus.

\begin{example}\label{ex:hardwareassignment}
	Consider the block diagram of Example~\ref{ex:connectedft}, and the E/E-architecture as in Example~\ref{ex:eearchitecture}.
	The mapping of the blocks $h(B_1) = \text{ECU}_1$, $h(B_2) = h(B_3) = \text{ADAS}_1$ induces a unique mapping of the channels according to the scheme described above: $h((B_1, B_3)) = \text{CAN-BUS}$ and $h((B_2, B_3)) = \internal{\text{ADAS}_1}$.
\end{example}

\subsubsection{Constructing a complete fault tree}
\label{Subsec:Constructing_complete_FT}
To obtain a complete DFT of the vehicle guidance system, we take the disjoint union of the  DFT $F$ for the system and block layer, and the hardware DFTs $\{ F_y \mid y \in \hardwareplatforms \cup \buses \}$.
The DFT $F$ contains annotations $Y^B$ for each block $B$.
For each $B \in \blocks$, an \FDEP with trigger $\toplevel{F_{\hardwareassignment(B)}}$ and dependent event $Y^B$ is added, and an \ADEP in the reverse direction.
The \FDEP ensures that a failure of the hardware platform causes the function blocks executed on that hardware to fail.
 The \ADEP ensures that the hardware FTs are correctly activated if the corresponding function blocks are activated.
Furthermore, for each channel $c = (B,B') \in \channels$, an \FDEP with trigger $\toplevel{F_{\hardwareassignment(c)}}$ and dependent event $I^{B'}(c)$ is added. 
The FDEP ensures that if the bus fails, the input of the target block fails as well.

\begin{figure}
\centering
\scalebox{0.750}{
    \begin{tikzpicture}
        \definecolor{hw}{RGB}{150,0,0}
        \colorlet{orig}{darkblue}
        \colorlet{connec}{black}
 	
	\node[or2,orig] (b1) at (1,1.5) {};
	\node[labelbox,orig] (b1_label) at (b1.east) {$B_1$};
	
    \node[or2,orig] (b2) at (3,1.5) {};
    \node[labelbox,orig] (b2_label) at (b2.east) {$B_2$};
	
    \node[or3,orig] (b3) at (6,1.5) {};
	\node[labelbox,orig] (b3_label) at (b3.east) {$B_3$};
	
	\node[or2,orig,below=1.5cm of b3, yshift=1.6cm] (or2) {};
	\node[labelbox,orig] (in_label) at (or2.east) {input};
	
	\node[be,draw=orig,below=1cm of b3, xshift=0.4cm] (hw) {};
	\node[labelbox,orig] (hw_label) at (hw.north) {hw};
	
	\node[be,draw=orig, below=1.4cm of or2.center, xshift=-0.8cm] (D) {};
	\node[labelbox,orig] (d_label) at (D.north) {in 2};
	\node[be,draw=orig, below=1.4cm of or2.center, xshift=0.8cm] (C) {};
	\node[labelbox,orig] (c_label) at (C.north) {in 1};

    \node[be,draw=orig] (1a) at (0.5, 0.1) {};
	\node[labelbox,orig] (1a_label) at (1a.north) {hw};
	\node[be,draw=orig] (1b) at (1.5, 0.1) {};
	\node[labelbox,orig] (1b_label) at (1b.north) {intern};
	\node[be,draw=orig] (2a) at (3.5, 0.1) {};
	\node[labelbox,orig] (2a_label) at (2a.north) {hw};
	\node[be,draw=orig] (2b) at (2.5, 0.1) {};
	\node[labelbox,orig] (2b_label) at (2b.north) {intern};
	
	\draw[-,orig] (b1.input 1) -- (1a_label.north);
	\draw[-,orig] (b1.input 2) -- (1b_label.north);
	\draw[-,orig] (b2.input 2) -- (2a_label.north);
	\draw[-,orig] (b2.input 1) -- (2b_label.north);

	\draw[-,orig] (b3.input 1) -- (in_label.north);
	\draw[-,orig] (b3.input 3) -- (hw_label.north);
	\draw[-,orig] (or2.input 1) -- (d_label.north);
	\draw[-,orig] (or2.input 2) -- (c_label.north);
	
	\node[or3,hw] (adas) at (10,1) {};
	\node[labelbox,hw] (adas_label) at (adas.east) {ADAS$_1$};
	
	\node[or3,hw] (ecu) at (-1.5,1) {};
	\node[labelbox,hw] (ecu_label) at (ecu.east) {ECU$_1$};
	
	\node[fdep,connec, scale=0.8] (f3) at (-0.5, 3.5) {};
	
	\draw[-,connec] (f3.T) -- (ecu_label.north);
	\draw[-,connec] (f3.ED) -- (1a_label.north);
	
	\node[fdep,connec, scale=0.8, xscale=-1] (f4) at (4.5, 3.5) {};

	\draw[-,connec](f4.T) -- (adas_label.north);
	\draw[-,connec](f4.ED) -- (2a_label.north);
	\draw[-,connec](f4.EA) -- (hw_label.north);
	\draw[-,connec](f4.EB) -- (d_label.north);
	
	\node[fdep,connec, scale=0.8, xscale=-1] (f5) at (8.2, 0.3) {};

	\node[or3,hw] (can) at (10,-1.5) {};
	\node[labelbox,hw] (can_label) at (can.east) {CAN-BUS};
	
	\draw[-,connec] (f5.T) -- (can_label.north);
	\draw[-,connec] (f5.EB) -- (c_label.north);
					
\end{tikzpicture}}
\caption{Fragment of a complete fault tree for a toy example}
\label{fig:excomplete}	
\end{figure}
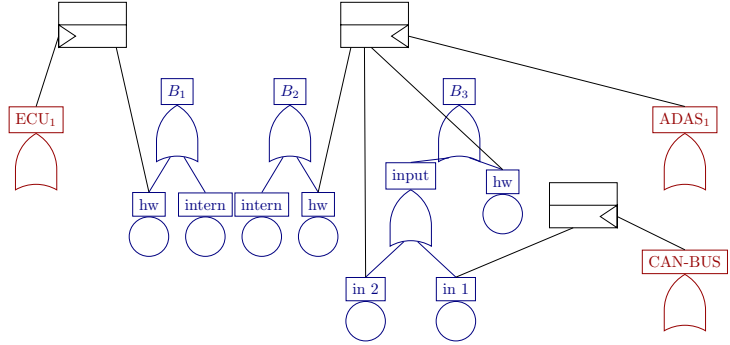

\begin{example}
We connect the system and block layer from Example~\ref{ex:functionalft} with hardware fault trees.
The connection is based on the hardware assignment in Example~\ref{ex:hardwareassignment}.
The relevant fragment of the complete fault tree is depicted in Fig.~\ref{fig:excomplete}.
Blue colour indicates the block fault trees and red colour the hardware fault trees.
\end{example}

\section{Analysing the Dynamic Fault Trees}
\label{sec:analysis}
Our aim is to analyse the complete FTs constructed as in Sect.~\ref{sec:modelling} with respect to the eight measures described in Sect.~\ref{Sec:VehicleGuidance_Properties}.
We build upon the model checker \storm{}~\cite{DJKV17}, a state-of-the-art tool for the automated analysis of DFTs~\cite{VJK17}.

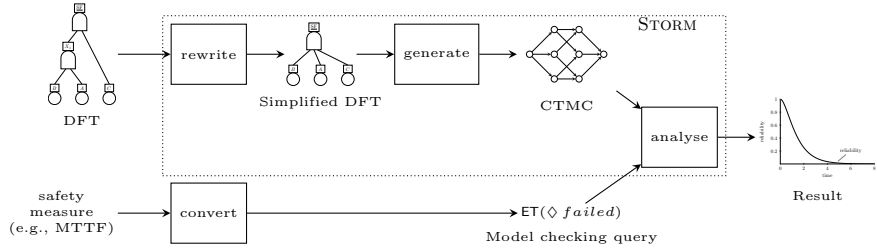
\begin{figure}[t]
	\tikzstyle{block} = [draw, rectangle, minimum height=0.8cm, minimum width=1.0cm]%
\centering
\begin{tikzpicture}[box/.style={draw, densely dotted, inner sep=5pt}]
\tikzstyle{every node}=[font=\tiny]
 \node[block] (rewrite_box) {rewrite};
 \node[left=0.7cm of rewrite_box, scale=0.24, minimum width=2cm, label=below:DFT] (in_dft) {\includegraphics{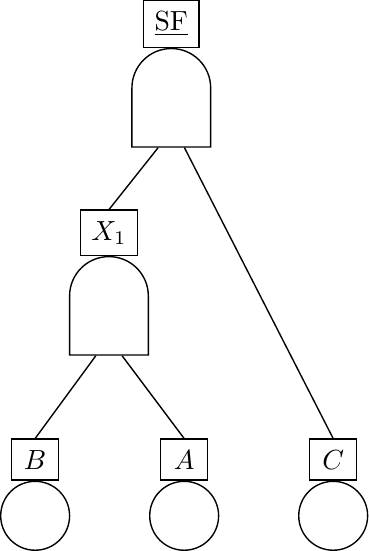}};
 \node[right=0.5cm of rewrite_box, scale=0.24, minimum width=3.2cm, label=below:Simplified DFT] (rewrite_dft) {\includegraphics{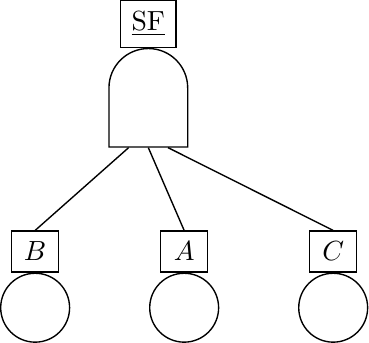}};
 \node[block, right=0.5cm of rewrite_dft] (generate_box) {generate};
 \node[right=0.5cm of generate_box,  minimum width=1.2cm, label=below:CTMC] (ctmc) {\scalebox{0.24}{\begin{tikzpicture}[line width=0.05cm]
     \node[state] (s1) {};
     \node[state, right=of s1] (s2) {};
     \node[state, above=of s2] (s3) {};
     \node[state, below=of s2] (s4) {};
     \node[state, right=of s2] (s5) {};
     \node[state, above=of s5] (s6) {};
     \node[state, below=of s5] (s7) {};
     \node[state, right=of s5] (s8) {};

     \draw[->] (s1) -- (s2);
     \draw[->] (s1) -- (s3);
     \draw[->] (s1) -- (s4);
     \draw[->] (s2) -- (s6);
     \draw[->] (s2) -- (s7);
     \draw[->] (s3) -- (s5);
     \draw[->] (s3) -- (s6);
     \draw[->] (s4) -- (s5);
     \draw[->] (s4) -- (s7);
     \draw[->] (s5) -- (s8);
     \draw[->] (s6) -- (s8);
     \draw[->] (s7) -- (s8);
  \end{tikzpicture}}};
 \node[block, below=0.2cm of ctmc, xshift=1.5cm] (analyse_box) {analyse};
 \node[right=0.5cm of analyse_box, scale=0.3, minimum width=4.6cm, label=below:Result] (result_graph) {\includegraphics{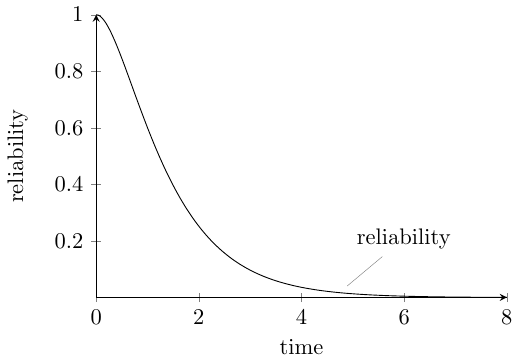}};
 \node[box, fit=(rewrite_box) (analyse_box), inner sep=3pt, label={[label distance=-1.1cm]17: \scriptsize{\storm}}] (storm) {};

 \node[block, below=1.3cm of rewrite_box] (translate_box) {convert};
 \node[left=0.7cm of translate_box, inner sep=1pt, align=center] (measure) {safety\\ measure\\(\eg \MTTF)};
 \node[right=3.6cm of translate_box, inner sep=1pt, label=below:Model checking query] (mc_query) {$\ET(\finally{failed})$};

 \draw[->] (in_dft.east) -- (rewrite_box);
 \draw[->] (rewrite_box) -- (rewrite_dft.west);
 \draw[->] (rewrite_dft.east) -- (generate_box);
 \draw[->] (generate_box) -- (ctmc);
 \draw[->] (ctmc) -- (analyse_box);
 \draw[->] (measure) -- (translate_box);
 \draw[->] (translate_box) -- (mc_query);
 \draw[->] (mc_query) -- (analyse_box);
 \draw[->] (analyse_box) -- (result_graph);
\end{tikzpicture}%
\caption{Overview of the DFT analysis}
\label{fig:dft_toolchain}
\end{figure}

The full tool-chain for the analysis is depicted in Fig.~\ref{fig:dft_toolchain}.
First, the DFT is simplified, and converted into a CTMC.
Details are given in Sect.~\ref{sec:analysis:modelgeneration}.
Second, a measure is translated into a model-checking query, as detailed in Sect.~\ref{Subsec:Translating_measures}.
Then, the result for the query on the CTMC is computed based on traditional probabilistic model checking, as detailed in Sect.~\ref{sec:analysis:mc}.

\subsection{Model generation}\label{sec:analysis:modelgeneration}
In the following we describe the translation from a DFT into a CTMC.
This translation is a key step in the analysis process.
The main challenge is the possible state-space explosion.
Several techniques such as rewriting DFTs~\cite{JGKRS17}, or partial state-space generation~\cite{VJK17} have been developed to address the issue.

\subsubsection{Rewriting DFTs}
The structured creation of FTs induces an extensive structure, which is often counterproductive for the analysis performance.
Thus, the first step is to \emph{rewrite} the given FT using the techniques from~\cite{JGKRS17}.
Rewriting  aims to automatically simplify the FT to improve the performance of its analysis while preserving the semantics of the original FT.
Examples of rewriting include the elimination of superfluous levels of \OR{s}, and the removal of \FDEP{s} whose trigger already lead to the failure of the top-level event.

\subsubsection{State space generation}
Next, the simplified FT is translated into a CTMC~\cite{VJK17, Boudali2010} by a process referred to as state-space generation.
Transitions in the CTMC correspond to \BE{s} failing.
States record in which order \BE{s} have failed and administer the status of children of \SPARE{s}, \eg are they currently in use or not.
Additionally, states are labelled to denote which events have failed.

The state-space generation is computationally the most expensive step as it explores the complete state space defined by the successive failures of \BE{s}.
To improve the performance, several optimisations are used~\cite{VJK17}.
For example, \storm{} exploits symmetric structures occurring in the FT modelling the sensors.

When using the default translation, the FT is converted into a Markov Automaton~\cite{EHZ10b}, an extension of CTMCs with non-determinism.
The non-determinism stems from different orders in which \FDEP{s} propagate their failure.
However, in our setting, the order of \FDEP{} failures does not influence the obtained analysis result.
Therefore, we fix a failure order for dependent events and restrict ourselves to MAs without non-determinism, that is, CTMCs.

\subsubsection{Transient faults}
Transient faults are considered during the state-space generation.
Recall from Sect.~\ref{sec:transientfaultassumption}, that if transient fault occurs, either the system fails directly, or the fault disappears and the system returns to its previous state.
During the state-space generation transient faults are considered in each state similar to regular \BE{} faults.
However, the corresponding transition is only added to the CTMC if the TLE has failed at the target state of the particular transition.

\subsection{Converting measures}
\label{Subsec:Translating_measures}
The generated CTMCs are analysed against properties defined in continuous stochastic logic (CSL) with reward extensions~\cite{BHHHK13}.
We convert the safety measures introduced in Sect.~\ref{Sec:VehicleGuidance_Properties} into model-checking queries composed of standard CSL properties.

\subsubsection{CSL properties}
Sets of states in a CTMC are described by Boolean combinations over the state labelings.
Thus, for a CTMC constructed from a fault tree, sets of states can be described by combinations of failed and operational events.
The following three standard CSL properties can be solved efficiently and are the building blocks for our queries.

\paragraph{Reach-avoid probability}
Given a state $s \in \states$, a set of \target states and a set of \bad states, the property $\PROB^s(\until{\bad}{\target})$ describes the probability to eventually reach a \target state from state $s$ without visiting a \bad state in-between.
If the set of \bad states is empty, the reach-avoid probability reduces to $\PROB^s(\finally{\target})$, and is just a \emph{reachability} probability.
If $s$ is the initial state, then we omit the superscript and write $\PROB(\finally{\target})$.

\paragraph{Time-bounded reach-avoid probability}
Incorporating an additional time-bound $t$, the property $\PROB^s(\boundeduntil{\bad}{t}{\target})$ describes the probability to reach a \target state (while avoiding \bad states) from state $s$ within time-bound $t$. Similar as before the \emph{time-bounded reachability} is described by $\PROB^s(\boundedfinally{t}{\target})$.

\paragraph{Expected time}
Given state $s$ and a set of \target states, the property $\ET^s(\finally{\target})$ describes the expected time to reach a \target state from $s$. The expected time is only defined if the reachability probability to reach \target is one.

\subsubsection{Conversion}
The conversion of the measures in Sect.~\ref{Sec:VehicleGuidance_Properties} into model-checking queries is given in Tab.~\ref{Table:CSL}.
\begin{table}[tb]
\centering
\scalebox{0.9}{
\renewcommand{\arraystretch}{1.3}
\begin{tabular}{c|c|c}
& Measure & Model-checking queries\\
\hline
\multirow{3}{*}{\rotatebox[origin=c]{90}{System}}
& Reliability & $1 - \PROB(\boundedfinally{t}{\text{failed}})$ \\
& \AFH & $\frac{1}{ \text{lifetime}} \cdot \PROB(\boundedfinally{\text{lifetime}}{\text{failed}})$ \\
& \MTTF & $\ET(\finally{\text{failed}})$ \\
\hline
\multirow{5}{*}{\rotatebox[origin=c]{90}{Degradation}}
& \FFA & $1 - \PROB\big(\boundedfinally{t}{(\text{failed} \vee \text{degraded})}\big)$ \\
& \FWD & $\PROB\big(\boundeduntil{(\neg \text{degraded})}{t}{(\neg \text{degraded} \wedge \text{failed})}\big)$ \\
& \MTDF & $\Sigma_{s \in \text{degraded}} \big(\PROB(\until{\neg \text{degraded}}{s}) \cdot \ET^s(\finally{\text{failed}})\big)$ \\
& \MDR & $\text{argmin}_{s \in \text{degraded}} \big(1 - \PROB^s(\boundedfinally{t}{\text{failed}})\big)$ \\
& \FLOD & $\Sigma_{s \in \text{degraded}} \big(\PROB(\boundeduntil{\neg \text{degraded}}{t}{s}) \cdot \PROB^s(\boundedfinally{\text{drivecycle}}{\text{failed}})\big)$ \\
& \SILFO & $1 - \Big(\text{\FWD} + \text{\FLOD}\Big)$ \\
\end{tabular}
}
\caption{Model-checking queries}
\label{Table:CSL}
\end{table}
The atomic proposition $failed$ denotes states where the top-level event in the FT has failed and $degraded$ denotes states where only reduced functionality is available.

The first three model-checking queries consider the safety-performance of the complete system.
For \emph{reliability}, the time-bounded reachability of a failed state is considered.
The  complement of this probability describes the probability that the system is still operational within time-bound $t$.
The \emph{average failure-probability per hour (\AFH)} is obtained similarly by using the lifetime of the system as time-bound and afterwards normalising by the lifetime.
\emph{Mean time to failure (\MTTF)} is the expected time of failure of the top-level event.
In the considered DFTs, the expected time is always defined.

The second set of model-checking queries describes the influence of degradation in the system.
\begin{compactenum}[1)]
    \item \emph{Full Function Availability (\FFA)} describes the time-bounded probability that the system provides full functionality, \ie it is neither failed nor degraded.
It is described as the complement of the time-bounded reachability of a failed or degraded state.
    \item \emph{Failure Without Degradation (\FWD)} describes the time-bounded probability that the system fails without being degraded first.
It is the time-bounded reach-avoid probability of reaching a failed state without reaching a degraded state.
    \item \emph{Mean Time from Degradation to Failure (\MTDF)} describes the expected time from the moment of degradation to system failure.
It is obtained by taking the expected time of failure for each degraded state and scaling it with the probability to reach this state while not being degraded before.
    \item \emph{Minimal Degraded Reliability (\MDR)} describes the criticality of degraded states by giving the worst-case failure probability when using the system in a degraded state.
For all degraded states the time-bounded reachability of a TLE failure is computed. The \MDR is the minimum over the complement of this result for all degraded states.
    \item \emph{System Integrity under Limited Fail-Operation (\SILFO)} considers the system-wide impact of limiting the degraded operation time.
\SILFO is split into two parts considering failures without degradation (\FWD) and failures with degradation (\FLOD).
\emph{Failure under Limited Operation in Degradation (\FLOD)} describes the probability of failure when imposing a time limit for using a degraded system.
For all degraded states the time-bounded reachability probability of a failed state is computed within the restricted time-bound given by a drive cycle.
This value is scaled by the time-bounded reach-avoid probability of reaching a degraded state without degradation before.
\end{compactenum}
For sensitivity analysis, several model-checking queries on CTMCs of DFTs with different failure rates are performed. 

\subsection{Analysis via model checking}
\label{sec:analysis:mc}
The resulting CTMC is analysed \wrt the given model-checking queries by using state-of-the-art probabilistic model checkers such as \storm{}~\cite{DJKV17}.

\subsubsection{Model checking}
The first five measures are computed with a single model-checking query each.
Furthermore, it is possible to compute (time-bounded) reach-avoid probabilities starting in a state $s$ for all states $s \in \states$ simultaneously.
Thus, \MDR can be obtained by computing $\PROB^s(\boundedfinally{t}{failed})$ for all degraded states in a single query and afterwards computing the minimum over the complement of all results.

However, the computation of \MTDF and \SILFO requires the computation of (time-bounded) reach-avoid probabilities reaching a degraded state $s$ for all degraded states $s \in \states$ independently.
A naive implementation therefore would require a model-checking query for each degraded state, increasing the computation time drastically.
Using similar ideas as in~\cite{KKNP01}, we implemented an improved algorithm computing (time-bounded) reach-avoid probabilities for all states simultaneously.
The improved algorithm works as follows.

The standard approach for reach-avoid probabilities is a ``backwards'' computation of the probabilities starting from the \target states.
The result is a vector of the (time-bounded) probabilities to reach a \target state for each state.
The idea here now is to perform a ``forwards'' computation of the probabilities starting from the initial state.
The result is a vector of (time-bounded) probabilities to reach each \target state from the initial state.
The improved algorithm performs a model-checking query only once for all degraded states at the same time.
Thus, the computation of \MTDF can be reduced to performing one model-checking query for $\PROB(\until{\neg degraded}{s})$, one query for $\ET^s(\finally{failed})$ and combining the results afterwards.
The computation of \SILFO can be performed similarly by combining the results of the three model-checking queries -- for \FWD, $\PROB(\boundeduntil{\neg degraded}{t}{s})$ and $\PROB^s(\boundedfinally{drivecycle}{failed})$.

\subsubsection{Evidence}
To flexibly support the analysis of degrades states, we use the concept of \emph{evidence}.
Evidence is given as a set of \BE{s} considered as already failed and all possible failure orderings (traces) of these \BE{s} are examined.
Following the traces from the initial state yields a set $S'$ of states describing the system status based on the given evidence.
Computing a model-checking query using evidence gives results starting in each state $s \in S'$.
When the complete state space has been built before, it is beneficial to reuse it for analysing the degraded system.

\subsubsection{Approximation}
\begin{figure}[t]
\centering
\scalebox{0.8}{
\begin{tikzpicture}
 	\node[rectangle, draw, align=center] (ft) {DFT};
 	\node[rectangle, draw, align=center, below=1.7cm of ft] (part) {Partial\\state space};
 	\node[circle, fill, inner sep=2pt, right=1.7cm of part] (partp) {};

 	\node[rectangle, draw, align=center, right=0.25cm of partp, yshift=0.6cm, minimum width=2.5cm] (OverMA) {CTMC\\upper bound};
 	\node[rectangle, draw, align=center, right=0.25cm of partp, yshift=-0.6cm, minimum width=2.5cm] (UnderMA) {CTMC\\lower bound};
 	\node[rectangle, draw, align=center, right=2cm of UnderMA, yshift=0.6cm] (approx) {Approximation\\$[l, u]$};
 	\node[rectangle, draw, align=center, above=1.9cm of approx] (res) {Result};
    \node[below=1.8cm of part] (dot) {};

 	\draw[->] (part) -- node[above] {extend} (partp);
 	\draw[->] (partp) -- (OverMA);
 	\draw[->] (partp) -- (UnderMA);
	\draw[<-] (approx) -- node[above, align=center] {model\\checking} (OverMA);
 	\draw[<-] (approx) -- node[below, align=center] {model\\checking} (UnderMA);

 	\draw[->] (ft) -- node[right, align=center] (otf) {partial\\generation} (part);
 	\draw[->, dashed] (approx) -- node[left] {contains} (res);
 	\draw[->] (approx) -- +(0, -1.5) -| node[below, pos=0.3] (ref) {refinement} (part);

 	\node[fit=(otf)(ref)(approx)(part)(dot), draw, dotted] (approxbox) {};
 	\node[below=-0.3cm of approxbox.south east, anchor=east, align=center, fill=white] {On-the-fly approximation};
\end{tikzpicture}
}
\caption{Approximation, based on~\cite{VJK17}}
\label{Fig:Approximation}
\end{figure}
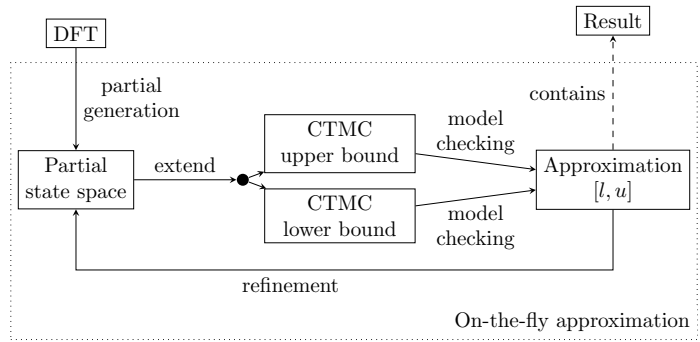

The main bottleneck of the DFT analysis is the state-space explosion problem.
For \MTTF computations, the problem can be alleviated by building only the most relevant fraction of the state space and derive approximative results~\cite{VJK17}.
The idea is visualised in Fig.~\ref{Fig:Approximation}.
For a given DFT a partial state space is generated.
From this partial state space two CTMCs are built.
One CTMC captures the behaviour over-approximating the exact result and one CTMC under-approximates the result.
The former is obtained by assuming that all remaining (\ie still operational) \BE{s} have to fail in order for the DFT to fail, the latter results by assuming that a failure of any single \BE{} leads to a failure.
By model checking both CTMCs an approximation $[l, u]$ can be derived.
The exact result for the original DFT is guaranteed to lie in this interval.
If the interval $[l, u]$ is too coarse, the state space is refined by exploring additional parts of the state space.
Incremental refinement of the state space approximates the exact result up to the desired precision.

We extend the approximation algorithm of~\cite{VJK17} to also compute the reliability.
To over-approximate the probability of a system failure we assume that each terminal state in the partial state space corresponds to the failure of the TLE.
The CTMC for the under-approximation is constructed by assuming that each terminal state is absorbing.
Due to the absorption, the TLE cannot fail in the future if it has not failed in one of the terminal states.

\section{Experiments}
\label{sec:experiments}
We show the applicability of the proposed methodology on systems and concepts similar to those from Sect.~\ref{Sec:VehicleGuidance_caee}.
The generated (anonymized) DFTs are available on our website\footnote{\url{http://www.stormchecker.org/publications/dfts-for-vehicle-guidance-systems}}.

\subsection{Set-up}
All experiments are executed on a 2.9 GHz Intel Core i5 with 8 GB RAM and a time-out (TO) of one hour.
We consider the three safety concepts SC1, SC2 and SC3.
For each safety concept, we construct a system- and block layer fault tree as described in Sect.~\ref{Subsec:FBD2FFT}.
All scenarios include four sensors, of which at least two are required for safe operation, and four actuators, which are all required for safe operation.

\subsubsection{Different partitioning schemes}
We discuss several partitioning schemes for the safety concepts, which we refer to as scenarios.
Each scenario is defined by the safety concept, the used architecture with possible adaptions, the hardware assignment,  and the fraction of sensors and actuators which have to be operational.
An overview of (a selection of) considered scenarios is given in Tab.~\ref{Table:Scenario}.
We consider E/E-architectures A, B and C depicted in Fig.~\ref{fig:ee},  and vary the architectures, by e.g.\ considering a redundant bus or introducing more ADAS platforms.
Additionally we consider partitions which do not use the I-ECU, and instead assign functions to the ADAS$_2$.
Furthermore, we scale the number of sensors and actuators and the required number of sensors for safe operation.

\begin{table}[tb]
\centering
\caption{Scenarios}
\scalebox{0.9}{
\begin{tabular}{c|c|c|c|r|r}
Scenario & \multicolumn{1}{c|}{Safety concept} & \multicolumn{1}{c|}{Architecture} & \multicolumn{1}{c|}{Adaptions} & \multicolumn{1}{c|}{Sensors} & \multicolumn{1}{c}{Actuators} \\
\hline
I    & SC1 & B & |            & 2/4 & 4/4 \\
II   & SC2 & B & |            & 2/4 & 4/4 \\
III  & SC2 & C & ADAS+        & 2/4 & 4/4 \\
IV   & SC3 & C & |            & 2/4 & 4/4 \\
V    & SC2 & A & |            & 2/4 & 4/4 \\
VI   & SC2 & B & removed I-ECU& 2/4 & 4/4 \\
VII  & SC2 & B & 5 ADAS, 2 BUS& 2/8 & 7/7 \\
VIII & SC2 & B & 8 ADAS, 2 BUS& 2/8 & 7/7 \\
\end{tabular}
}
\label{Table:Scenario}
\end{table}

\subsubsection{Failure rates}
\label{sec:failure_rates}
For presentation purposes we assume the following failure rates, which do not necessarily reflect reality and especially do not reflect any system from BMW AG.
We assume that function blocks, \eg EP, are free of systematic (internal) faults.
Sensors, actuators, and (I-)ECUs have failure rates of $10^{-7}/h$.
In the ADAS hardware platforms transient faults occur with rate $10^{-4}/h$ and permanent faults occur with rate $10^{-5}/h$.
All faults can be detected by a safety mechanism which itself fails with rate $10^{-5}/h$.
The safety mechanism for ASIL D covers 99\% of the faults, the safety mechanism for ASIL B covers 90\%, and the one for ASIL QM covers 60\% of the faults.
For the ADAS+ platform, the failure rates increase by a factor $10$ and the safety mechanism covers 99.9\% of all faults.

\subsubsection{Tool-support}
The complete workflow depicted in Fig.~\ref{fig:overview}, \ie both the generation of the DFTs as well as their analysis, are supported by a Python tool-chain.
Given a safety concept as a function block diagram and the FT for each block, the tool-chain first automatically generates a FT where dependencies are inserted according to the data flow in the safety concept.
Given an E/E architecture with a partitioning and hardware FTs, and the system- and block layer FT generated before, the complete FT is automatically constructed.
This generation of the complete FT is performed in milliseconds.
Finally, the analysis of the complete FT as described in Fig.~\ref{fig:dft_toolchain} is performed fully automatically as well.
Timings for the analysis are given in Tab.~\ref{Table:Timing}.

We describe the failure rates in the hardware FTs symbolically, i.e., as parameters.
Thus, changes in coverage or failure rates require only a different instantiation of the parameters, instead of reconstructing the FT.

\subsection{Evaluation}
In the following we consider the scenarios of Tab.~\ref{Table:Scenario}.
The characteristics of the corresponding DFTs and CTMCs can be found in Tab.~\ref{Table:Model}.
\begin{table}[tb]
\centering
\caption{Model characteristics}
\scalebox{0.9}{
\begin{tabular}{c|r|r|r|r|r|r}
 & \multicolumn{3}{c|}{DFT} & \multicolumn{3}{c}{CTMC}\\
Scen. & \multicolumn{1}{c|}{\#BE} & \multicolumn{1}{c|}{\#Dyn.} & \multicolumn{1}{c|}{\#Elem.} & \multicolumn{1}{c|}{\#States} & \multicolumn{1}{c|}{\#Trans.} & \multicolumn{1}{c}{Degrad.}\\
\hline
I    & 76 &25 &233 &    5,377 &    42,753 &     |\\
II   & 70 &23 &211 &    5,953 &    50,049 & 19\%\\
III  & 57 &19 &168 &    1,153 &     7,681 & 17\%\\
IV   & 57 &21 &170 &      385 &     1,985 & 12\%\\
V    & 58 &19 &185 &      193 &       897 &  0\%\\
VI   & 65 &21 &199 &    1,201 &     8,241 & 20\%\\
VII  & 96 &30 &266 &  109,369 & 1,148,785 & 19\%\\
VIII &114 &36 &305 &5,179,105 &84,454,945 & 11\%\\
\end{tabular}
}
\label{Table:Model}
\end{table}
The first column identifies the scenario.
The following three columns give the number of \BE{s}, dynamic gates and the total number of nodes in the DFT.
The last three columns describe the CTMC obtained after generating the state space and applying reduction techniques.
The columns contain the number of states and transitions, and the percentage of degraded states in the CTMC.

The analysis results for the measures from Sect.~\ref{Subsec:Translating_measures} are given in Tab.~\ref{Table:Measures}.
Notice that SC1 does not contain degraded states and in scenario V the system fails before reaching a degradation.
We use a lifetime of 10,000 hours, and a drive cycle of 1 hour~\cite[5:9.4]{iso26262}.
\begin{table}[t]
\centering
\setlength{\tabcolsep}{5pt}
\caption{Obtained measures with operational lifetime=10,000h and drive cycle=1h ($x^c$ indicates complement $1-x$)}
\scalebox{0.9}{
\begin{tabular}{c|r|r|r|r|r|r|r|r|r}
 & \multicolumn{3}{c|}{System} & \multicolumn{6}{c}{Degradation}\\
Scen.& \multicolumn{1}{|c|}{$\text{rel.}^c$} & \multicolumn{1}{c|}{\AFH} & \multicolumn{1}{c|}{\MTTF} & \multicolumn{1}{c|}{$\text{\FFA}^c$} & \multicolumn{1}{c|}{\FWD} & \multicolumn{1}{c|}{\MTDF} & \multicolumn{1}{c|}{\MDR} & \multicolumn{1}{c|}{\FLOD} & \multicolumn{1}{c}{$\text{\SILFO}^c$}\\
\hline
I    &1.6E-2 &1.6E-6 &8.6E+4 &\multicolumn{1}{c|}{|} &\multicolumn{1}{c|}{|} &\multicolumn{1}{c|}{|} &\multicolumn{1}{c|}{|} &\multicolumn{1}{c|}{|} &\multicolumn{1}{c}{|}\\
II   &1.0E-2 &1.0E-6 &3.4E+5 &5.2E-2 &1.0E-2 &2.3E+5 &2.9E-1 & 4.3E-8 &1.0E-2\\
III  &1.2E-2 &1.2E-6 &1.1E+5 &5.2E-2 &1.1E-2 &2.1E+4 &7.4E-1 & 2.7E-7 &1.1E-2\\
IV   &1.0E-2 &1.0E-6 &3.1E+5 &1.6E-2 &1.0E-2 &1.6E+5 &2.1E-1 & 6.5E-9 &1.0E-2\\
V    &6.0E-2 &6.0E-6 &6.9E+4 &6.0E-2 &6.0E-2 &     0 &0      & 0      &6.0E-2\\
VI   &1.1E-2 &1.1E-6 &3.4E+5 &5.3E-2 &1.1E-2 &2.3E+5 &2.1E-1 & 4.7E-8 &1.1E-2\\
VII  &1.7E-2 &1.7E-6 &2.8E+5 &5.8E-2 &1.7E-2 &1.7E+5 &3.7E-1 & 7.2E-8 &1.7E-2\\
VIII &1.7E-2 &1.7E-6 &2.7E+5 &9.8E-2 &1.6E-2 &2.0E+5 &4.3E-1 & 1.4E-7 &1.6E-2\\
\end{tabular}
}
\label{Table:Measures}
\end{table}
The times for generating the CTMC from the DFT and computing each measure on the CTMC are given in Tab.~\ref{Table:Timing}.
\begin{table}[t]
\centering
\caption{Timings}
\scalebox{0.9}{
\begin{tabular}{c|r|r|r|r|r|r|r|r}
 & \multicolumn{1}{c|}{I} & \multicolumn{1}{c|}{II} & \multicolumn{1}{c|}{III} & \multicolumn{1}{c|}{IV} & \multicolumn{1}{c|}{V} & \multicolumn{1}{c|}{VI} & \multicolumn{1}{c|}{VII} & \multicolumn{1}{c}{VIII} \\
\hline
CTMC generation    & 0.52s                  & 0.51s & 0.08s & 0.05s & 0.02s & 0.10s & 12.02s &2043.78s \\
$\text{reliability}^c$  & 0.03s             & 0.03s & 0.00s & 0.00s & 0.00s & 0.00s &  1.00s &  82.84s \\
\AFH               & 0.03s                  & 0.04s & 0.01s & 0.00s & 0.00s & 0.01s &  0.93s & 157.94s \\
\MTTF              & 0.01s                  & 0.01s & 0.00s & 0.00s & 0.00s & 0.00s &  0.18s &  26.43s \\
$\text{\FFA}^c$    & \multicolumn{1}{c|}{|} & 0.02s & 0.01s & 0.00s & 0.00s & 0.00s &  0.54s &  64.41s \\
\FWD               & \multicolumn{1}{c|}{|} & 0.02s & 0.00s & 0.00s & 0.00s & 0.00s &  1.46s &  61.79s \\
\MTDF              & \multicolumn{1}{c|}{|} & 0.48s & 0.10s & 0.02s & 0.02s & 0.10s & 10.78s & 826.73s \\
\MDR               & \multicolumn{1}{c|}{|} & 0.49s & 0.09s & 0.02s & 0.03s & 0.09s & 11.02s & 829.27s \\
\FLOD              & \multicolumn{1}{c|}{|} & 1.08s & 0.20s & 0.05s & 0.04s & 0.21s & 28.49s &2945.81s \\
$\text{\SILFO}^c$  & \multicolumn{1}{c|}{|} & 1.10s & 0.20s & 0.05s & 0.04s & 0.21s & 29.95s &3007.60s \\
\end{tabular}
}
\label{Table:Timing}
\end{table}

Fig.~\ref{Fig:Results} illustrates the obtained measures for a variety of concepts and architectures.
\pgfplotsset{every axis/.append style={
                    label style={font=\scriptsize},
                    tick label style={font=\tiny}  
                    }}%
\begin{figure}[tb]
\centering
\subfigure[Probability of failure]{
\begin{minipage}{0.47\textwidth}
\centering
  \begin{tikzpicture}
    \begin{axis}[
        width=4.7cm,
        height=4.7cm,
        xmin=0, ymin=0, xmax=50000, ymax=0.3,
        axis x line=bottom, axis y line=left,
        x label style={at={(axis description cs:0.5,0.06)},anchor=north},
        y label style={at={(axis description cs:0.16,0.5)},anchor=south},
        xlabel=Time (h), ylabel=Failure probability,
        scaled ticks=false,
        xtick={0, 25000, 50000}, xticklabels={0, 25k, 50k},
        yticklabel style={font=\tiny}, xticklabel style={rotate=0, anchor=north, font=\tiny},
        mark repeat={25},
        legend pos=north west, legend style={font=\tiny}]
            \addplot[color=orange,mark=x]
        table [col sep=semicolon, x=time, y=safety_concept_1_b_normal]
        {tikz_figures/timepoints_all.csv};
      \addplot[color=blue,mark=*]
        table [col sep=semicolon, x=time, y=safety_concept_2a_b_normal]
        {tikz_figures/timepoints_all.csv};
      \addplot[color=red,mark=square*]
        table [col sep=semicolon, x=time, y=safety_concept_2b_b_normal]
        {tikz_figures/timepoints_all.csv};
      \addplot[color=green,mark=triangle*]
        table [col sep=semicolon, x=time, y=safety_concept_3_b_normal]
        {tikz_figures/timepoints_all.csv};
      \legend{I, II, III, IV}
    \end{axis}
  \end{tikzpicture}
\end{minipage}
\label{Fig:Integrity_all}
}
\hfill
\subfigure[Average failure-prob. per hour]{
\begin{minipage}{0.47\textwidth}
\centering
  \begin{tikzpicture}
    \begin{axis}[
        width=4.7cm,
        height=4.7cm,
        xmin=1, ymin=0, xmax=50000, ymax=0.0000054,
        axis x line=bottom, axis y line=left,
        x label style={at={(axis description cs:0.5,0.06)},anchor=north},
        y label style={at={(axis description cs:0.16,0.5)},anchor=south},
        xlabel=Life time (h), ylabel=\AFH,
        scaled ticks=false,
        xtick={0, 25000, 50000}, xticklabels={0, 25k, 50k},
        ytick={0, 0.000001, 0.000002, 0.000003, 0.000004, 0.000005}, yticklabels={0, 1e-6, 2e-6, 3e-6, 4e-6, 5e-6},
        yticklabel style={font=\tiny}, xticklabel style={rotate=0, anchor=north, font=\tiny},
        mark repeat={25},
        legend pos=north west, legend style={font=\tiny}]
      \addplot[color=orange,mark=x]
        table [col sep=semicolon, x=time, y=safety_concept_1_b_normal-der]
        {tikz_figures/timepoints_all.csv};
      \addplot[color=blue,mark=*]
        table [col sep=semicolon, x=time, y=safety_concept_2a_b_normal-der]
        {tikz_figures/timepoints_all.csv};
      \addplot[color=red,mark=square*]
        table [col sep=semicolon, x=time, y=safety_concept_2b_b_normal-der]
        {tikz_figures/timepoints_all.csv};
      \addplot[color=green,mark=triangle*]
        table [col sep=semicolon, x=time, y=safety_concept_3_b_normal-der]
        {tikz_figures/timepoints_all.csv};
      \legend{I, II, III, IV}

    \end{axis}
  \end{tikzpicture}
\end{minipage}
\label{Fig:AFH}
}
\subfigure[Sensitivity analysis]{
\begin{minipage}{0.47\textwidth}
\centering
  \begin{tikzpicture}
    \begin{axis}[
        width=4.7cm,
        height=4.7cm,
        xmin=0, ymin=0, xmax=50000, ymax=0.35,
        axis x line=bottom, axis y line=left,
        x label style={at={(axis description cs:0.5,0.06)},anchor=north},
        y label style={at={(axis description cs:0.16,0.5)},anchor=south},
        xlabel=Time (h), ylabel=Failure Probability,
        scaled ticks=false,
        xtick={0, 25000, 50000}, xticklabels={0, 25k, 50k},
        ytick={0, 0.1, 0.2, 0.3}, yticklabels={0, 0.1, 0.2, 0.3},
        yticklabel style={font=\tiny}, xticklabel style={rotate=0, anchor=north, font=\tiny},
        mark repeat={25},
        legend pos=north west, legend style={font=\tiny}]
                                                                                                                  \addplot[color=red,mark=square*]
        table [col sep=semicolon, x=time, y=safety_concept_2b_b_normal]
        {tikz_figures/timepoints_all.csv};
      \addplot[color=red,mark=square*,mark options={solid},dashed]
        table [col sep=semicolon, x=time, y=safety_concept_2b_b_high]
        {tikz_figures/timepoints_all.csv};
      \addplot[color=red,mark=square*,mark options={solid},dotted]
        table [col sep=semicolon, x=time, y=safety_concept_2b_b_low]
        {tikz_figures/timepoints_all.csv};
      \addplot[color=green,mark=triangle*]
        table [col sep=semicolon, x=time, y=safety_concept_3_b_normal]
        {tikz_figures/timepoints_all.csv};
      \addplot[color=green,mark=triangle*,mark options={solid},dashed]
        table [col sep=semicolon, x=time, y=safety_concept_3_b_high]
        {tikz_figures/timepoints_all.csv};
      \addplot[color=green,mark=triangle*,mark options={solid},dotted]
        table [col sep=semicolon, x=time, y=safety_concept_3_b_low]
        {tikz_figures/timepoints_all.csv};
      \legend{III,inc.,dec., IV,inc.,dec.}
    \end{axis}
  \end{tikzpicture}
\end{minipage}
\label{Fig:Sensitivity}
}
\subfigure[SILFO]{
\begin{minipage}{0.47\textwidth}
\centering
  \begin{tikzpicture}
    \begin{axis}[
        width=4.7cm,
        height=4.7cm,
        xmin=0, ymin=0, xmax=50000, ymax=0.11,
        axis x line=bottom, axis y line=left,
        x label style={at={(axis description cs:0.5,0.06)},anchor=north},
        y label style={at={(axis description cs:0.16,0.5)},anchor=south},
        xlabel=Time (h), ylabel=Failure probability,
        scaled ticks=false,
        xtick={0, 25000, 50000}, xticklabels={0, 25k, 50k},
        ytick={0, 0.05, 0.1}, yticklabels={0, 0.05, 0.1},
        yticklabel style={font=\tiny}, xticklabel style={rotate=0, anchor=north, font=\tiny},
        mark repeat={2},
        legend pos=north west, legend style={font=\tiny}]
      \addplot[color=blue,mark=*]
        table [col sep=semicolon, x=time, y=safety_concept_2a_b_normal]
        {tikz_figures/timepoints_silfo.csv};
      \addplot[color=red,mark=square*]
        table [col sep=semicolon, x=time, y=safety_concept_2b_b_normal]
        {tikz_figures/timepoints_silfo.csv};
      \addplot[color=green,mark=triangle*]
        table [col sep=semicolon, x=time, y=safety_concept_3_b_normal]
        {tikz_figures/timepoints_silfo.csv};
      \legend{II, III, IV}
    \end{axis}
  \end{tikzpicture}
\end{minipage}
\label{Fig:Degradation}
}
\caption{Analysis results}
\label{Fig:Results}
\end{figure}
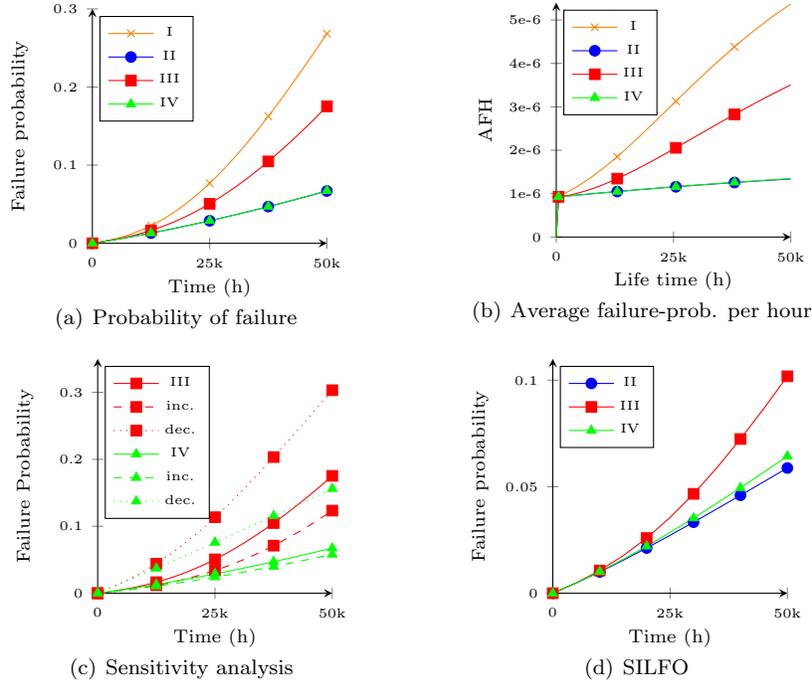%

In Fig.~\ref{Fig:Integrity_all} the complement of the reliability, \ie the failure probability, over a lifetime of 50,000 hours is given for the scenarios I-IV.
Fig.~\ref{Fig:AFH} depicts the \AFH.
Fig.~\ref{Fig:Sensitivity} compares the sensitivity of the failure probability for scenarios III and IV.
For the sensitivity analysis we change the ASIL levels of the hardware FTs, \ie change the coverage of the safety mechanism.
In both scenarios the straight lines are obtained by the baseline failure rates and coverages for the hardware components as given in Sect.~\ref{sec:failure_rates}.
The dashed (dotted) lines are obtained assuming an increased (decreased) coverage according to an increased (decreased) ASIL level, respectively.
The graph in Fig.~\ref{Fig:Degradation} displays the SILFO for the safety concepts with degraded states.

Results for the approximation are given in Fig.~\ref{Fig:Approx_Results}.
\pgfplotsset{every axis/.append style={
                    label style={font=\scriptsize},
                    tick label style={font=\tiny}  
                    }}%
\begin{figure}[tb]
\centering
\subfigure[Failure probability for VIII]{
\begin{minipage}{0.47\textwidth}
\centering
  \begin{tikzpicture}
    \begin{axis}[
        width=4.7cm,
        height=4.7cm,
        xmode=log,
        xmin=0, ymin=0.015, xmax=100, ymax=0.02,
        axis x line=bottom, axis y line=left,
        x label style={at={(axis description cs:0.5,0.06)},anchor=north},
        y label style={at={(axis description cs:0.16,0.5)},anchor=south},
        xlabel=Time (s), ylabel=Failure probability,
        scaled ticks=false,
        xtick={1, 10, 100}, xticklabels={1s, 10s, 100s},
        ytick={0.015, 0.017, 0.02}, yticklabels={1.5e-2, 1.7e-2, 2.0e-2},
        yticklabel style={font=\tiny}, xticklabel style={rotate=0, anchor=north, font=\tiny},
        mark repeat={1},
        legend pos=south east, legend style={font=\tiny}]
      \addplot[color=blue,mark=*]
        table [col sep=semicolon, x=Time, y=ResultUnder]
        {tikz_figures/approx_2g.csv};
      \addplot[color=red,mark=square*]
        table [col sep=semicolon, x=Time, y=ResultOver]
        {tikz_figures/approx_2g.csv};
      \legend{Lower bound, Upper bound}
    \end{axis}
  \end{tikzpicture}
\end{minipage}
\label{Fig:Approx_Integrity}
}
\subfigure[Approximation error]{
\begin{minipage}{0.47\textwidth}
\centering
  \begin{tikzpicture}
    \begin{axis}[
        width=4.7cm,
        height=4.7cm,
        xmode=log,
        ymode=log,
        xmin=0, ymin=0, xmax=1000, ymax=2,
        axis x line=bottom, axis y line=left,
        x label style={at={(axis description cs:0.5,0.06)},anchor=north},
        y label style={at={(axis description cs:0.16,0.5)},anchor=south},
        xlabel=Time (s), ylabel=Error bound,
        scaled ticks=false,
        xtick={1, 10, 100, 1000}, xticklabels={1s, 10s, 100s, 1000s},
        ytick={1e-9, 1e-5, 1e-1}, yticklabels={1e-9, 1e-5, 1e-1},
        yticklabel style={font=\tiny}, xticklabel style={rotate=0, anchor=north, font=\tiny},
        mark repeat={1},
        legend pos=south west, legend style={font=\tiny}]
      \addplot[color=blue,mark=*]
        table [col sep=semicolon, x=Time, y=Error]
        {tikz_figures/approx_2e.csv};
      \addplot[color=red,mark=square*]
        table [col sep=semicolon, x=Time, y=Error]
        {tikz_figures/approx_2g.csv};
      \legend{VII, VIII}
    \end{axis}
  \end{tikzpicture}
\end{minipage}
\label{Fig:Approx_Error}
}
\caption{Approximation results}
\label{Fig:Approx_Results}
\end{figure}
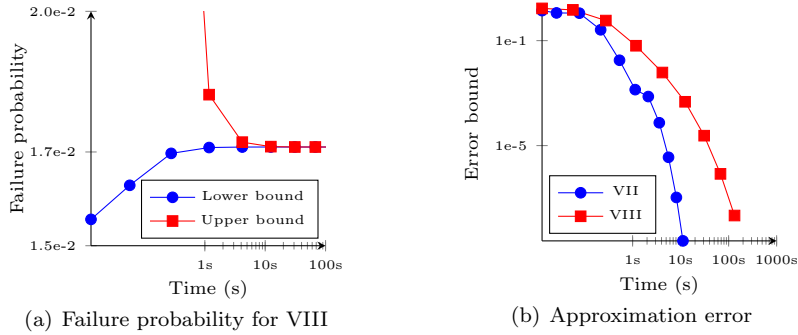%

Fig.~\ref{Fig:Approx_Integrity} illustrates the lower and upper bound for the failure probability in scenario VIII \wrt the computation time.
Fig.~\ref{Fig:Approx_Error} depicts the approximation error over the computation time for the largest scenarios.
\emph{Moreover, computing an approximate reliability allowing a 1\% relative error on scenario VIII requires only 206,410 states and could be computed within 12 seconds.}

\section{Discussion}
\label{sec:discussion}
\subsection{Analysis of results}
We evaluate the obtained results \wrt the assumed failure rates.
The evaluation is not intended as a recommendation for a specific partitioning scheme, but they illustrate the possibility to perform design space exploration efficiently.
In the following we exemplarily evaluate the results of the scenarios I-IV.
The variety of measures computed allows some insights in the effect of different safety concepts and the role of degradation.

The \AFH and \MTTF indicate that system reliability of scenarios II and IV are superior compared to III and I.
The differences between II and IV are marginal, and III is better than I.
These claims can also be deduced from Fig.~\ref{Fig:Integrity_all} and Fig.~\ref{Fig:AFH}.
The similarity between I and III stems from the fact that in both cases the system fails if two paths fail, \ie two out of three in the TMR of I, or normal and safety path in III.
It is interesting to see that the encoding in ADAS+ for III only marginally improves the reliability, because the better fault coverage of the encoding is outweighed by the higher load on the hardware.
In scenario II, all three paths---one normal and two redundant safety paths---have to fail before the complete system fails.
In scenario IV, the fallback path has a reduced hardware load as long as the primary path is still operational, leading to a smaller failure rate for this path.

However, from the sensitivity analysis in Fig.~\ref{Fig:Sensitivity} we can deduce that the lessons are only valid with respect to the assumed failure rates.
In particular, for scenario IV increasing the safety coverage only has a marginal effect on the failure probability.
Thus, the benefit of increasing fault coverage in platforms depends on the chosen architecture.

Scenarios II and IV differ in their failure behaviour of degraded states as seen in Fig.~\ref{Fig:Degradation}.
When limiting the driving time in the degraded state to one hour, scenario II offers a better reliability than IV whereas in the overall reliability the difference is marginal.
The \FLOD is orders of magnitude smaller than \FWD in all scenarios.
Thus, the duration of the drive cycle is insignificant for \SILFO.

The results in Fig.~\ref{Fig:Approx_Results} show that we can obtain tight approximation results for the standard measures within seconds, even for the largest scenarios.

\subsection{General methodology}
We discuss the methodology based on generating DFTs for function block diagrams and hardware assignments.

\paragraph{Direct translation to CTMCs}
A direct translation from the system description to CTMCs is arguably more flexible, and allows to keep any overhead to a minimum.
However, even a naive translation is necessarily complex and error-prone, and the resulting CTMCs are typically too large to be comprehensible.
It is hard to give the modeller feedback on the meaning of the constructed CTMC.
Fault trees, in contrast, are comparably small, and contain more structure.
Additionally, state-space generations have to be implemented with performance in mind, which makes the direct translation likely to be error prone.

Moreover, an additional benefit of reusing the DFT formalism is due to the presence of tool-support.
The state-space generation only had to be slightly adapted, which is significantly easier than the construction of an efficient state-space generation from scratch.

\paragraph{Automation of fault tree generation}
\label{sec:manual_creation}
Manual creation of the system- and block layer of the fault tree has some advantages, noteworthy are:
\begin{compactitem}
	\item The semantics of the function block diagram do not need to be formalised. In particular, function block diagrams contain implicit assumptions, e.g., assumptions about different failure behaviour of voters, or channels with different meanings. Manual creation can adapt for these subtle differences.
	\item  Constructing the FT is an important step in the development-\emph{process} of safety-critical systems~\cite{iso26262}.
\end{compactitem}
Due to the structure, as discussed in Sect.~\ref{sec:modelling}, our prototype made several default suggestions, \eg for block FTs, that greatly reduce the required manual effort.

The generation of the complete fault tree given the system and block layers is fully automated.
The automation is essential for the proposed methodology, as this allows a push-button comparison of the various possible variants of hardware assignments.

\paragraph{Using dynamic fault trees}
Using DFTs as the underlying model has several advantages.
Fault trees are a well-known concept in reliability engineering.
Their hierarchical structure allows for a faithful model of the different layers in the considered scenarios.
Using DFTs instead of static fault trees provides more expressiveness.
For example, most of the proposed measures for degraded states cannot be computed on static fault trees.
The proposed fault trees contain several features only present in DFTs, including the gates \PAND (for switches), \SPARE (for cold redundancy), \SEQ{} (in safety mechanisms).
Functional dependencies are heavily used, in particular to simplify the representation of feedback loops.
The claiming mechanism of \SPARE gates is not used. 
The claiming mechanism has traditionally led to some strong separation assumptions of the subtrees under \SPARE{s}, and does restrict some possibilities for simplification.
DFTs traditionally lack activation dependencies, but they could be added straightforwardly. 
In particular, they do not seem more complex than the existing gates or dependencies.
Thus, most features of DFTs are present in the generated fault trees and DFTs with the addition of activation dependencies seem a suitable formalism to assess the failure behaviour.

\paragraph{Analysis methods}
Fig.~\ref{Fig:AFH} indicates that the average failure-probability per hour (\AFH) varies for different operation life times.
This observation justifies the analysis \wrt different measures and time horizons.
Tab.~\ref{Table:Model} indicates that reduction techniques successfully alleviate the state-space problem.
The generated state space remains small even for hundreds of elements.
The size of the state space depends largely on the scenario.
Naturally, latent faults increase the state space, but then the effectiveness of the approximation increases as shown in Fig.~\ref{Fig:Approx_Results}.
Using CTMCs as an underlying model allows to check a wide variety of measures out-of-the-box.
Tab.~\ref{Table:Timing} indicates that most of the measures can be computed within seconds even on the largest models.
The more complex measures as \MTDF and \SILFO require a tailored implementation to avoid performing model-checking queries for each degraded state.
However, the tailored implementation was able to reuse the existing building blocks of the model checker \storm.
The approximation algorithm computes tight results for the reliability and \MTTF within seconds.
Thus, the approximation scales well for large scenarios with millions of states which can be analysed quickly by only building the most relevant fraction of the state space.

\subsection{Related work}
Earlier work~\cite{MS09} considers an automotive case study where functional blocks are translated to static fault trees without treating the partitioning on hardware architectures.
\cite{KolblL18} has a similar setting but focuses more on causal explanations and less on analysis performance of large-scale models.
The evaluation of various options from the design space by a translation to fault trees, and applying fault tree analysis has also been considered for air traffic control~\cite{DBLP:conf/cav/GarioCMTR16}.
The effect of different topologies of a FlexRay bus has been assessed using FTA in  \cite{LCWC12}; and identified the need for modelling dynamic aspects.
The analysis of architecture decisions under safety aspects has been considered in e.g.~\cite{RBFAKS14} using a dedicated description language and an analytical evaluation.
Safety analysis for component-based systems has been considered in~\cite{GKP05}, using state-event fault trees.
Qualitative FTA has been used in~\cite{AOMM13} for ISO 26262 compliant evaluation of hardware.
Different hardware partitionings are constructed and analysed using an Architecture Description Language (ADL) in~\cite{WWRPPLMCS13}.
ADL-based dependability analysis has been investigated for several languages, e.g.,
AADL~\cite{BCKNNR11}, UML~\cite{LL11}, Arcade~\cite{BCHKS08}, and HiP-HOPS~\cite{DBLP:conf/safecomp/ChenJLPSTT08}.
These approaches typically have a steeper learning curve than the use of DFTs.
The powerful M\"obius analysis tool~\cite{CGKRS09}  has recently been extended with dynamic reliability blocks~\cite{DBLP:journals/sesa/KeefeS16}.
Model checking for safety analysis has been proposed by, e.g., \cite{DBLP:journals/scp/BozzanoCLMMRT15}; which focuses on AltaRica, and does not cover probabilistic aspects.

DFTs are a subclass of the more expressive state/event fault trees (SEFTs) \cite{DBLP:journals/ress/KaiserGF07}, but efficient analysis techniques for SEFTs are lacking.
Various variants and analysis techniques for DFTs exist~\cite{JGKS16}.
A precise comparison for the semantics of state-based approaches for DFT analysis is given in~\cite{JKSV18}.
Model checking for DFT was first proposed by~\cite{Boudali2010}. The performance of that approach suffers from an intrinsic overhead.
Static/Dynamic fault trees~\cite{DBLP:conf/safecomp/BackstromBHKK16} are a subclass of DFTs, and allow for efficient analysis, but lack the expressive power to express ordered failure and warm redundancy.
To support a richer class of failure distributions and improve scalability, rare event simulation for Fault-Maintenance Trees~\cite{DBLP:conf/qest/RuijtersGDPS16} has been considered \cite{DBLP:conf/safecomp/RuijtersRBS17}.

\subsection{Future Work}
\noindent Future work can be partitioned into two directions:
\begin{compactitem}
\item \emph{improved modelling}, either by improved expressive power, or by conciseness.
\item \emph{improved analysis}, either by the support for a broader range of properties, or by improved performance for existing properties.
\end{compactitem}

Regarding modelling, we would like to investigate the modelling of involved error propagation schemes, possible containing deterministic timing information.
A possible direction would be to consider a combination with timed failure propagation graphs~\cite{DBLP:conf/ijcai/BittnerBC16}.
Similar combinations have been considered within the COMPASS project~\cite{DBLP:journals/scp/BozzanoCLMMRT15}.
It would be interesting to consider the use of Boolean driven Markov decision processes~\cite{DBLP:journals/ress/BouissouB03} or SEFTs~\cite{DBLP:journals/ress/KaiserGF07}.

For an improved analysis, a more rigorous treatment of transient faults and the combination with degraded states seems promising, for which we may draw inspiration from~\cite{DBLP:conf/qest/RuijtersGDPS16}.
Moreover, for an improved failure rate sensitivity analysis, we would like to investigate the use of parametric Markov models~\cite{DBLP:conf/tacas/CeskaPPBK16,DBLP:conf/atva/QuatmannD0JK16}.

\section{Conclusion}
\label{sec:conclusions}
We presented a model-based approach using dynamic fault trees towards the safety analysis of
vehicle guidance systems.
The approach (see~Fig.~\ref{fig:overview}) takes the system functions and their mapping onto the hardware architecture into account.
Its main benefit is the flexibility: new partitionings and architectural changes can easily and automatically be accommodated.
The use of DFTs instead of static FTs allows for a more faithful model, \eg incorporating warm and cold redundancies, and order-dependent failures.
The obtained DFTs were analysed with probabilistic model checking.
Due to tailored state-space generation~\cite{VJK17} and reduction techniques, the analysis of these DFTs---with up to 100 basic events---is a matter of minutes.









\bibliographystyle{elsarticle-num}

\bibliography{bibliography}

\end{document}